\documentclass[letterpaper,10pt,nofootinbib,aps,tightenlines]{revtex4}

\usepackage{amsmath,amsfonts,amssymb}
\usepackage{mathrsfs}
\usepackage{graphicx}
\usepackage[english]{babel} 
\usepackage{hyperref} 
\usepackage{color}
\usepackage[section]{placeins}

\usepackage[cal=boondoxo]{mathalfa}

\usepackage[normalem]{ulem}

\usepackage[utf8]{inputenc}

\hypersetup{
    colorlinks=true,
    linkcolor=red,
    citecolor=blue,
} 

\newcommand{\gy}{g_{Y}}
\newcommand{\thau}{\tilde\tau}

\begin{document}

\title{Gravitational Wave Opacity from Gauge Field Dark Energy}
\author{R.~R.~Caldwell}
\author{C.~Devulder}
\affiliation{Department of Physics and Astronomy, Dartmouth College, 6127 Wilder Laboratory, Hanover, New Hampshire 03755 USA}

\date{\today}
%%%%%%%%%%%%%%%%%%%%%%%%%%%%%%%%%%%%%%%%%%%%%%%%%%%%%%%%%%%

%%%%%%%%%%%%%%%%%%%%%%%%%%%%%%%%%%%%%%%%%%%%%%%%%%%%%%%%%%%
\begin{abstract}
We show that astrophysical gravitational waves can undergo an anomalous modulation when propagating through cosmic gauge field dark energy. A sufficiently strong effect, dependent on the gauge field energy density, would appear as a redshift-dependent opacity, thereby impacting the use of gravitational wave standard sirens to constrain the expansion history of the Universe. We investigate a particular model of cosmic gauge field dark energy and show that at early times it behaves like dark radiation, whereas a novel interaction causes it to drive cosmic acceleration at late times. Joint constraints on the cosmological scenario due to type 1a supernovae, baryon acoustic oscillations, and cosmic microwave background data are presented. In view of these constraints, we show that standard siren luminosity distances in the redshift range $0.5 \lesssim z \lesssim 1.5$ would systematically dim by up to $1\%$, which may be distinguishable by third-generation gravitational wave detectors.

\end{abstract}
%%%%%%%%%%%%%%%%%%%%%%%%%%%%%%%%%%%%%%%%%%%%%%%%%%%%%%%%%%%
\maketitle

%%%%%%%%%%%%%%%%%%%%%%%%%%%%%%%%%%%%%%%%%%%%%%%%%%%%%%%%%%%
\section{Introduction}
\label{intro}

The recent detection of gravitational waves has led to the emergence of gravitational wave astronomy, opening a new vista to astrophysical phenomena. The tantalizing prospect of combining gravitational wave (GW) sources with an electromagnetic (EM) counterpart is expected to lead to the development of a new method to constrain the expansion history of the Universe \cite{Schutz:1986gp,Holz:2005df}. The GW profile of binary inspirals, for example, is so distinctive that the luminosity distance of the source can be inferred within $1-10\%$ uncertainty \cite{Hughes:2001ya}. Observatories such as advanced LIGO \cite{Harry:2010zz,TheLIGOScientific:2014jea}, the proposed Einstein Telescope \cite{Sathyaprakash:2009xt,Punturo:2010zz}, and the future space-based detector LISA \cite{Audley:2017drz} are expected to achieve the sensitivity required to make the detection of such ``standard sirens" commonplace, extending the reach of GW astronomy to high redshift.

The luminosity distance inferred from GW standard sirens is susceptible to novel effects that could be within reach of future GW detectors. In particular, we focus on the phenomenon of gravitational wave - gauge field (GWGF) oscillations, in which the amplitude of a GW modulates as it propagates through a cosmic gauge field \cite{Caldwell:2016sut}. In a dark energy model based on a homogeneous non-Abelian gauge field, this effect would result in a distinct imprint on astrophysical GWs. Specifically, GWs couple to wave-like excitations in a background gauge field. At high frequencies relevant for astrophysical sources, the system is akin to a pair of coupled oscillators: as energy exchange occurs between the two oscillators, the GW amplitude weakens and grows continuously, leading to temporal blind spots. An otherwise strong GW would thus arrive at our detectors with a much lower amplitude. To study this effect in a cosmological setting, we begin this article by considering a model of dark energy based on a SU(2) gauge field. While originally introduced in the context of primordial inflation \cite{Maleknejad:2011jw}, a similar model can also be used to address the present-day cosmic acceleration \cite{Mehrabi:2015lfa}. We show that the net effect on astrophysical GWs is a redshift-dependent reduction in amplitude which could be within reach of future GW observatories. In particular, we show that a network of third-generation detectors, such as the proposed Cosmic Explorer \cite{Evans:2016mbw} and Einstein Telescope \cite{Punturo:2010zz}, may be able to distinguish this novel effect.

The rest of the paper is organized as follows. In Sec.~\ref{sec2}, we present an illustrative model of GWGF oscillations. In Sec.~\ref{sec3}, we give an overview of gauge field dark energy, or gauge quintessence, as introduced in \cite{Mehrabi:2015lfa}, and investigate its properties as a dark energy component. We show that, in contrast with most models of dark energy, cosmic acceleration is temporary as the equation of state eventually returns to 1/3 in the future. We constrain the model using observational data from type 1a supernovae, baryon acoustic oscillations and cosmic microwave background experiments. In Sec.~\ref{sec4} we return to investigate the interplay between gravitational waves and the gauge field in a dark energy scenario. We present our main result, namely that gauge field dark energy can have a direct impact on the amplitude of astrophysical GWs. Our final discussion is in Sec.~\ref{sec5}. In Appendix~\ref{app1} we explain how to expand the gauge field dark energy to include multiple, disjoint SU(2) subgroups. In Appendix~\ref{app2} we show that the same phenomenon could seed a spectrum of primordial gravitational waves.

%%%%%%%%%%%%%%%%%%%%%%%%%%%%%%%%%%%%%%%%%%%%%%%%%%%%%%%%%%%
\section{A Simple Model}
\label{sec2}

To illustrate the effect of a cosmic gauge field on gravitational waves, we begin by considering a gauge field under general relativity:
\begin{equation}
S = \int d^4 x \sqrt{-g} \left( \frac{1}{2} M^2_P R - \frac{1}{4} F_{a\mu \nu} F^{a \mu \nu} + \mathcal{L}_m  \right)
\end{equation}
where we use metric signature $-+++$, $M_P$ is the reduced Planck mass, and $\mathcal{L}_m$ represents any other fields that may be present. We take an SU(2) field
\begin{equation}
    F^a_{\mu \nu} = \partial_\mu A^a_\nu - \partial_\nu A^a_\mu - \gy \epsilon^{abc} A^b_\mu A^c_\nu
    \label{eqn:fdefn}
\end{equation}
where $\gy$ is the Yang-Mills coupling. Greek letters will be used to represent spacetime indices, and Latin letters $i,j,...$ are used for spatial indices. To match the symmetries of our cosmological spacetime having line element $ds^2 = a^2(\tau)(-d\tau^2 + d \vec{x}^2)$, the gauge field is taken to be in a flavor-space locked configuration:
\begin{equation}
A^a_\mu = 
\begin{cases}
\phi(t) \delta^a_\mu & \mu = 1, 2, 3 \\
0 & \mu = 0
\end{cases}
\label{eqn:fsl}
\end{equation}
i.e. expressed in terms of a triad of mutually orthogonal vectors each associated with a direction of real space, as required to achieve an isotropic and homogeneous configuration.

\begin{figure}[b]
    \includegraphics[width=0.45\linewidth]{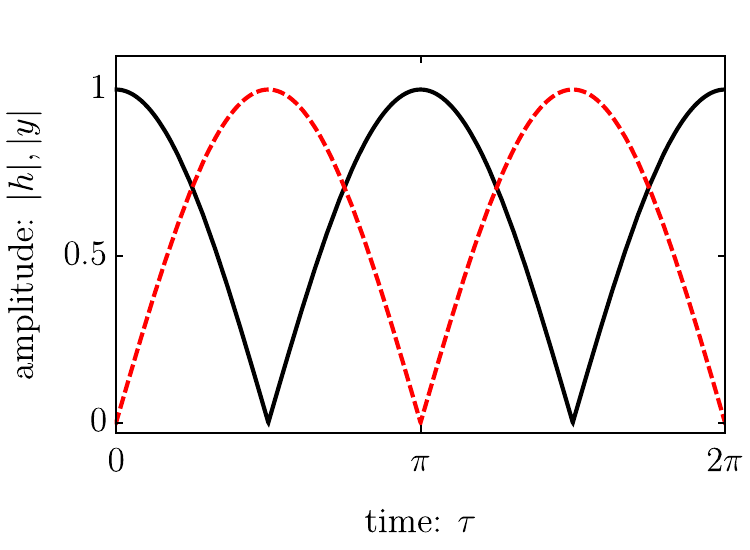}
    \hspace{1cm}
    \includegraphics[width=0.45\linewidth]{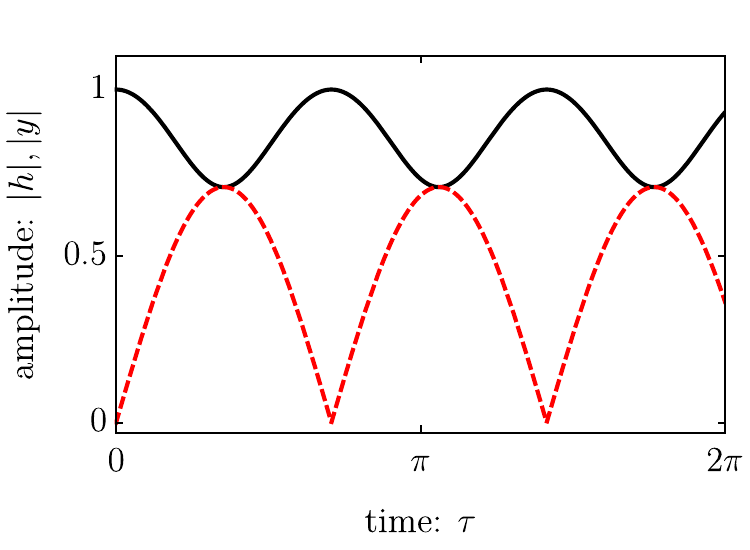}
\caption{(Left) The oscillations of the gravitational wave amplitude $\mathcal{h}$ (black) and gauge field $\mathcal{y}$ (dashed). The gravitational wave propagating through a stationary gauge field transforms into a gauge field wave and back again. Complete conversion of GW energy into the gauge field first occurs at $\tau_0 = \tau_{EMIT} + \pi/2$. (Right) Partial absorption of gravitational wave amplitude $\mathcal{h}$ by the gauge field $\mathcal{y}$.}
\label{fig:fig1.1}
\end{figure}

In this scenario, the gauge field fluctuates in the presence of a GW, in turn setting up travelling waves of its own. For ease of calculation, we consider a gravitational wave propagating in the $z$-direction:
\begin{equation}
\delta g_{\mu\nu} = a^2 h_{\mu\nu} = a^2 \left(\begin{array}{cccc}
0 & 0 & 0 & 0 \cr
0 & h_+ & h_\times & 0 \cr
0 & h_\times & -h_+ & 0 \cr
0 & 0 & 0 & 0 
\end{array}\right).
\label{eqn:Atnsr1}
\end{equation}
Similarly, we consider a $z$-directed gauge field wave
\begin{equation}
\delta A^a_\mu = a\,y^a_\mu = a \left( \begin{array}{cccc}
0 & y_+ & y_\times & 0 \cr
0 & y_\times & -y_+ & 0 \cr
0 & 0 & 0 & 0
\end{array}\right).
\label{eqn:Atnsr2}
\end{equation}
The equations of motion for left circularly polarized gravitational and gauge field waves decouple from the right circularly polarized modes.  However, the equations of motion are identical for both circular polarizations in the short-wavelength limit, for wavenumber $k$ much greater than both the expansion rate and the gauge field rate of change. Continuing, we write $H = a M_p h / \sqrt{2}$ and $Y= \sqrt{2} a y$, whereupon the Fourier amplitudes of monochromatic plane wave solutions to the equations of motion are then given by 
\begin{align}
H(\tau) & = e^{i b_3 \tau} \left[ c_0 \cos{b \tau} - \frac{1}{b} \sin{b \tau} \times ([b_1 + i b_2] c_\mathcal{1} + i b_3 c_0 ) \right] \times e^{-ik\tau} \\
Y(\tau) & = e^{i b_3 \tau} \left[ c_1 \cos{b \tau} + \frac{1}{b} \sin{b \tau} \left( [b_1 - i b_2] c_0 + i b_3 c_1 \right) \right] \times e^{-ik\tau}.
\label{eqn:Hsoln}
\end{align}
Here the initial values are $H(0) = c_0, Y(0) = c_1$, the wavenumber $k$ is taken to be large enough such that we can treat the coefficients $b_1 = \phi' /a M_P, b_2 = \gy \phi^2 / a M_P, b_3 = \gy \phi /2 $ as constants, and $b^2 = b_1^2 + b_2^2 +b_3^2$. The strength of the modulation effect is controlled by the coefficients $b_i$ which, as we show later, are determined by the gauge field energy density in units of the critical density. Note further that we measure conformal time in units of $H_0^{-1}$ and comoving wavenumbers $k,\, b$ in units of $H_0$. In the absence of the gauge field, the gravitational wave solution is simply $H = c_0 e^{ik\tau}$. Hence, the wave amplitude is constant, $c_0$, as we have already accounted for redshifting. In the presence of the gauge field, the GW amplitude modulates.

The interaction between gravitational and gauge field waves is revealed by their rms amplitude in the high frequency limit, for which we let $\mathcal{h}(\tau) = H(\tau) e^{ik\tau} $ and $\mathcal{y}(\tau) = Y(\tau) e^{ik\tau}$. Fig.~\ref{fig:fig1.1} shows the oscillations of $\mathcal{h}$ and $\mathcal{y}$ for the simple case $b_1 = 1, b_2 = b_3 = 0$. We can use these curves to determine the behavior of a GW burst emitted at a time $\tau_{EMIT}$ and observed at a time $\tau_{OBS}$. If the comoving separation is such that $\tau_{OBS} - \tau_{EMIT} = \pi/2$, then the GW will have completely converted into the gauge field by the present day. Hence, a gravitational wave observatory on Earth would record no signal. On the other hand, a GW burst from a more distant source whereby $\tau_{OBS} - \tau_{EMIT} = \pi$ would be unaffected.

In general, this energy conversion is only partial. The right panel illustrates the scenario where $b_1 = 0, b_2 = b_3 = 1$, resulting in a smaller fraction of the GW energy being converted into the gauge field. In this case, the maximum conversion occurs when $\tau_0 - \tau_{EMIT} =(2n+1) \pi/(2 \sqrt{2})$ for $n=1,\,2,\,\ldots $. Using standard cosmological parameters to compute conformal times, the most recent maximum conversion ($n=1$) corresponds to a redshift of emission of $z \simeq 1.7$ ($a/a_0 \simeq 0.4$).

\begin{figure}[t]
\includegraphics[width=0.45\textwidth,angle=0]{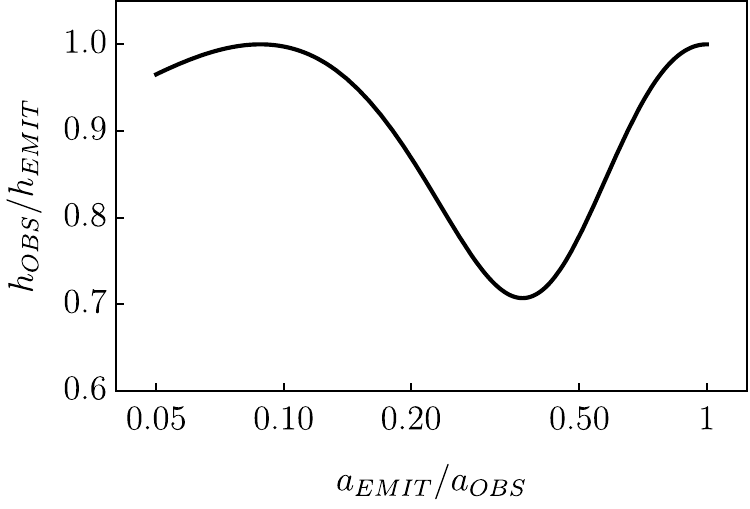}
\caption{Transmitted gravitational wave amplitude observed today as a function of the scale factor $a_\text{EMIT}$ at which it is emitted, for the case $b_1 = 0, b_2 = b_3 = 1$.}
\label{fig:fig1.2}
\end{figure}

To better illustrate this effect, we have plotted ${\cal h}(\tau_0 - \tau(a_{EMIT}))/{\cal h}_i$ versus scale factor $a_{EMIT}$ in Fig.~\ref{fig:fig1.2}, for the second scenario. This curve represents a sort of visibility function for gravitational waves. To interpret the curve, choose the scale factor at the time of emission along the horizontal axis, in which case the height of the curve gives the transmitted gravitational wave fraction. For example, a gravitational wave emitted at $a_{EMIT}/a_{OBS}\sim 0.3$ is dimmed by a factor $0.7$; a gravitational wave emitted at $a_{EMIT}/a_{OBS}\sim 0.1$ is not dimmed at all. It is easy to see that a standard siren cosmological distance determined from a GW amplitude will be systematically distorted by this effect. In order to make a realistic estimate of the distortion, next, we need to determine the properties of the gauge field in a viable cosmological scenario.

%%%%%%%%%%%%%%%%%%%%%%%%%%%%%%%%%%%%%%%%%%%%%%%%%%%%%%%%%%%
\section{Gauge Quintessence Model}
\label{sec3}

Gauge quintessence is a model of dark energy based on a non-Abelian gauge field \cite{Mehrabi:2015lfa} which provides a suitable cosmological framework to investigate GWGF interactions. The model consists of an SU(2) gauge field minimally coupled to gravity, as specified by the following action:
\begin{equation}
S = \int d^4 x \sqrt{-g} \left( \frac{1}{2} M^2_P R - \frac{1}{4}   F_{a \mu \nu} F^{a \mu \nu} + \frac{\lambda}{M^4_P} (F_{a \mu \nu} \tilde F^{a \mu \nu})^2   \right) + S_m(\chi_i, g_{\mu \nu} )
\label{eqn:action}
\end{equation}
where $R$ is the Ricci scalar and $S_m$ is the action for Standard Model particles. The dual field is $\tilde F^{a \mu \nu} = \frac{1}{2} \epsilon^{\mu \nu \alpha \beta} F^a_{\alpha \beta}$. The novel $\lambda$ term is responsible for the negative pressure, leading to cosmic acceleration.  More general actions based on an SU(2) gauge field have also been studied, e.g. Refs.~\cite{Allys:2016kbq,Rodriguez:2017wkg}. This theory is a toy model which we use in place of a more realistic theory that requires more parameters. In particular, we use this model  because cosmic acceleration driven by an axion field with Chern-Simons coupling, as in Chromo-Natural Inflation \cite{Adshead:2012kp,Adshead:2012qe}, can be equivalently described by Eq.~(\ref{eqn:action}). In this case, the parameter $\lambda$ can be related to the axion mass and the energy scale of the Chern-Simons coupling.

In the current scenario, the equation of motion for the gauge field follows from varying the action,
\begin{equation}
\nabla^\mu W^a_{\mu\nu}+\gy f^{abc} A_{b}^{\mu}W_{c\mu\nu}=0, \qquad
W^a_{\mu\nu} = F^a_{\mu\nu}-8\frac{\lambda}{M_P^4} (F\tilde{F})\tilde F^a_{\mu\nu}.
\label{eqn:coveom}
\end{equation}
The stress-energy tensor is  
\begin{equation}
T_{\mu \nu} = F_{a \mu \sigma}F^a_{\nu \tau} g^{\sigma \tau} - \frac{1}{4} g_{\mu \nu} F^2 - \frac{\lambda}{M_P^4} g_{\mu \nu} ( F \tilde F )^2.
\end{equation}
We consider cosmological solutions wherein directions of the internal SU(2) space are aligned with the principle axes of the Cartesian, spatially flat Robertson-Walker spacetime $ds^2 = a^2(\tau)(-d\tau^2 +   d \vec{x}^2)$, namely the ansatz known as flavor-space locking of Eqn.~(\ref{eqn:fsl}). For convenience we use primes to denote derivatives with respect to $\tau$, and define $\kappa \equiv 96 \lambda \gy^2$. The equation of motion for $\phi(\tau)$ then reads
\begin{equation}
\phi'' + 2\gy^2 \phi^3 + \frac{\kappa}{a^4 M_p^4} ( \phi^4 \phi'' + 2\phi^3 \phi'^2 - 4(a'/a) \phi^4 \phi'	) = 0.
\end{equation}
The energy density and pressure are given by
\begin{align}
\rho &= \frac{3}{2} \left( \frac{\phi'^2}{a^4} + \gy^2 \frac{\phi^4}{a^4} 	\right) + \frac{3}{2}\frac{\kappa}{M^4_P} \left( \frac{\phi' \phi^2}{a^4} \right)^2 
\label{eqn:rhophi}\\
p &=  \frac{1}{2} \left( \frac{\phi'^2}{a^4} + \gy^2 \frac{\phi^4}{a^4} 	\right) - \frac{3}{2}\frac{\kappa}{M^4_P} \left( \frac{\phi' \phi^2}{a^4} \right)^2
\label{eqn:pphi}
\end{align}
where the terms involving $\kappa$ are associated with dark energy, while the remaining contributions account for dark radiation. In this scenario, at early times the energy density is dominated by the $\kappa-$independent terms and behaves like a species of dark radiation. At early times, the approximate equation of motion $\phi'' + 2 \gy^2 \phi^3=0$ is solved by a cosine-like Jacobi elliptic function, $\phi = \phi_i {\rm sn}(\gy \phi_i \tau |-1)$. Although this equation of motion resembles that of a scalar field in a quartic potential, the story is slightly more complicated. Whereas the expression for the energy density may look familiar, the ``kinetic" term in the pressure has the wrong sign, such that the $\kappa-$independent equation of state is exactly $1/3$. As the field evolves monotonically and the product $\phi' \phi^2$ grows, the $\kappa$-dependent term dominates the energy density and pressure so that the equation of state approaches $w \to -1$, thereby driving cosmic acceleration. In this regime, the solution to the equation of motion is approximately $\phi \propto a(\tau)$ with $a \propto 1/(2 \tau_a-\tau)$ for de Sitter-like expansion. However, the equation of motion cannot sustain this solution indefinitely. As a consequence, acceleration is only temporary, for the rapid oscillations of $\phi$ eventually bring $w$ back up to $1/3$ in the future.

The coefficients $b_i$ that control the gravitational wave -- gauge field oscillations are now seen to be exactly the terms appearing in the radiation portion of the energy density in Eq.~(\ref{eqn:rhophi}). For example, $b_1 \propto a H \sqrt{\Omega_{E}}$ and $b_2 \propto a H \sqrt{\Omega_{B}}$ where $E,\,B$ refer to the electric and magnetic field-like contributions to the gauge field radiation energy density. The coefficient $b_3$ controls the oscillation frequency in the equation of motion $\phi'' + 2 \gy^2 \phi^3=0$. Hence, the modulation amplitude and frequency are directly related to the fraction of critical density in the gauge field radiation. Referring back to Eq.~(\ref{eqn:Hsoln}), we expect astrophysical gravitational waves to be modulated according to a factor
\begin{equation}
\times \left(1 - \frac{ b_1^2 + b_2^2}{2 b^2}\sin^2 b\tau \right)
\end{equation}
where $b_3 \gg b_1,\ b_2$. By the present day, we expect $b_3 \sim b_2 \Omega^{-1/4}$. Plugging in some rough numbers, for a radiation energy density contributing $1\%$ of critical density (meaning $w \simeq -0.99$) then the modulation amplitude is $\sim 5\%$. This sets our expectations for the magnitude of the gravitational wave opacity effect.

We have also considered the cosmological dynamics in the case that the SU(2) gauge field is replaced by SU(N), where $N$ is sufficiently large to contain ${\cal N}$ disjoint SU(2) subgroups. Details are provided in Appendix~\ref{app1}. There it is shown that the background equations for the case ${\cal N}>1$ are equivalent to the SU(2) case when the following identifications are made:
\begin{equation}
\phi_{\cal N} = \frac{1}{\sqrt{\cal N}}\phi_1, \qquad g_{\cal N} = \sqrt{\cal N} g_{1}, \qquad \kappa_{\cal N} = {\cal N} \kappa_{1}.
\end{equation}
Hence, our study is easily adapted to include larger gauge groups.

We further note that the phenomena of gravitational wave - gauge field oscillations leads to interesting effects at long wavelengths, too. In Refs.~\cite{Bielefeld:2014nza,Bielefeld:2015daa} we studied the effect of gauge field dark radiation on a primordial spectrum of long wavelength gravitational radiation. In Appendix~\ref{app2} we show how a spectrum of long wavelength tensor fluctuations of the gauge field can seed a similar spectrum of gravitational waves.

%%%%%%%%%%%%%%%%%%%%%%%%%%%%%%%%%%%%%%%%%%%%%%%%%%%%%%%%%%%
\subsection{E and B Solutions}

We present the results of the numerical solution of the background equations of motion. Initial values $\phi,\, \phi'$ and parameters $\gy,\,\kappa$ are chosen so that the dark energy equation of state is $w=1/3$ at early times, deep in the radiation era, and evolves $w \to -1$ by the present day. Since the dark energy component contributes a non-negligible fraction of the energy budget in the early stages of the matter dominated era, we refer to its contribution as ``early dark energy" (EDE). Here we illustrate the case where $\Omega_{EDE} = 0.001$ during the radiation era. The YM coupling sets the rate of oscillation of the gauge field, so that in order to enable slow roll-like behavior leading to cosmic acceleration we are compelled to set the coupling constant ${\hat g}_Y \equiv \gy M_P/H_0$ to order unity. One free parameter remains, $\kappa$, which we use to ensure the correct abundance of dark energy at the present day. The photon, neutrino, baryon, and dark matter densities are fixed according to the standard cosmology.

There are two classes of solutions of the gauge field quintessence equation of motion that describe a radiation-dominated fluid which later evolves into dark energy. These are the ``B" solutions, in which $\gy \phi^2 \gg |\phi'|$, and the ``E" solutions, in which $\gy \phi^2 \ll |\phi'|$. These two cases allow for the radiation-like portion of the energy density to dominate over the dark energy term. The nomenclature derives from the similarity with the electric and magnetic field components of the Faraday tensor of electromagnetism. The initial conditions for these two cases can be set as 
\begin{eqnarray}
\gy \phi_i^2 &=&  (2 \Omega_{r,0} R\cos^2\theta)^{1/2}  H_0 M_P a_i^2\\
\phi_i' &=&  (2 \Omega_{r,0} R\sin^2\theta)^{1/2} H_0 M_P a_i^2,
\end{eqnarray} 
where $R \equiv \Omega_{EDE}/\Omega_{r}$ and $\Omega_r$ is the fraction of critical density in photons and neutrinos. The above initial conditions ensure that the initial value of $\phi'^2_i + \gy^2\phi_i^4$, appearing in Eqs.~(\ref{eqn:rhophi}-\ref{eqn:pphi}), is a constant. Hence, the gauge field energy density and pressure start out scaling as $1/a^4$ and contribute a fraction $\Omega_{EDE}$ of the total energy density, deep in the radiation era.

\begin{figure}[htp]
\centering
\includegraphics[width=0.45\linewidth]{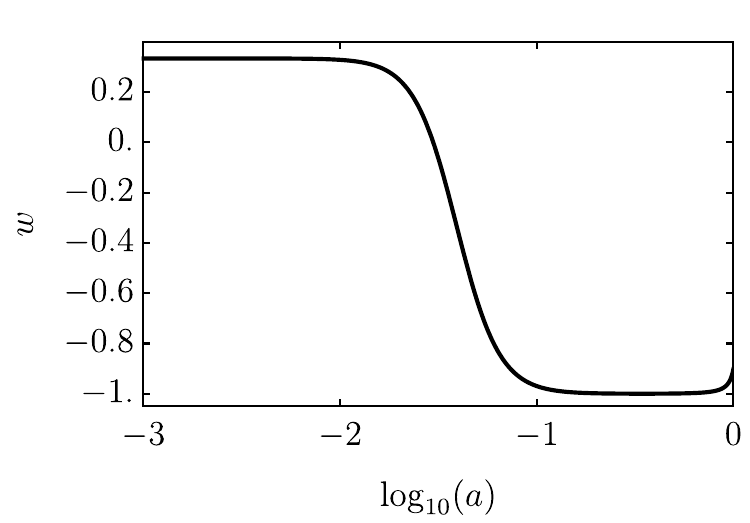}
\caption{Evolution of the equation of state $w(a)$ under the case of the ``B" solution. The time of transition to dark energy behavior is determined by $\gy$ and $\Omega_{EDE}$. For this model, ${\hat g}_{Y}=0.1,\, \kappa = 2.5\times 10^5,\, R=0.02,\, \Omega_M=0.3$ and the present day value of the equation of state is $w_0 = -0.9$. This model uses opposite signs for $\phi_i,\,\phi_i'$. Using like signs for the initial values $\phi_i,\,\phi_i'$, the equation of state curve is identical except with no upturn, such that $w_0=-0.99$.}
\label{fig:bsoln}
\end{figure}

The B solutions are obtained by setting $\theta \ll 1$. These are the solutions studied in Ref.~\cite{Mehrabi:2015lfa}. In this case, the equation of state is $w=1/3$ until late times, when it sharply steps down to $w \simeq -1$. The function $w(a) = (1 - 4 \tanh^n(a/a_*))/3$ provides a reasonable fit to the numerical solution, illustrated in Fig.~\ref{fig:bsoln}, where parameters $n\ge 1$ and $a_*$ depend on parameters $R$ and $\Omega_M$. We note that both signs ($\pm)$ of $\phi_i$ are considered; there is a slight difference in the solutions, but the broad features are the same. The background evolution of the gauge field quintessence closely resembles that of $\Lambda$CDM plus $\Delta N_\nu$ additional light neutrino species as dark radiation. As such, the model is as viable as one of the most basic extensions of the vanilla $\Lambda$CDM cosmology, as explored in Ref.~\cite{Mehrabi:2015lfa}. Current bounds $N_\nu = 3.15 \pm 0.23 (1 \sigma)$ \cite{Ade:2015xua} allow $\Delta N_\nu < 0.33$ and therefore $\Omega_{EDE} < 0.042 (1 \sigma)$. However, one interesting way in which the background evolution can differ from $\Lambda$CDM is suggested by the slight upturn in the equation of state shown in Fig.~\ref{fig:bsoln}. This upturn presages a sharp spike in $w$ up to $1/3$ in the future before returning back to near $-1$. By adjusting $\gy$ larger or smaller the spike can be moved to earlier times or later times, into the future, respectively. Cosmic acceleration eventually ends in this model, but only after many e-foldings of the accelerated cosmic expansion.

\begin{figure}[htp]
\centering
\hspace*{-5.5cm}
\includegraphics[width=0.4\linewidth]{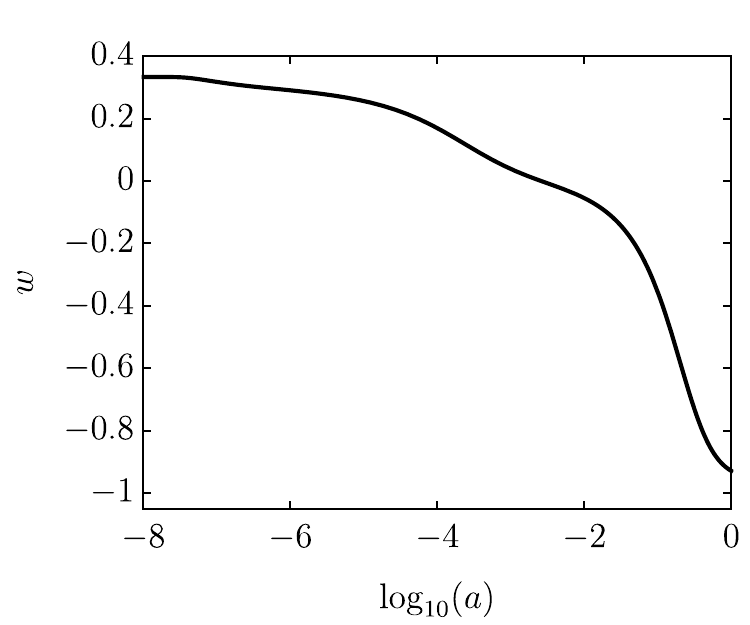}
\hspace*{0cm}
\includegraphics[width=0.4\linewidth]{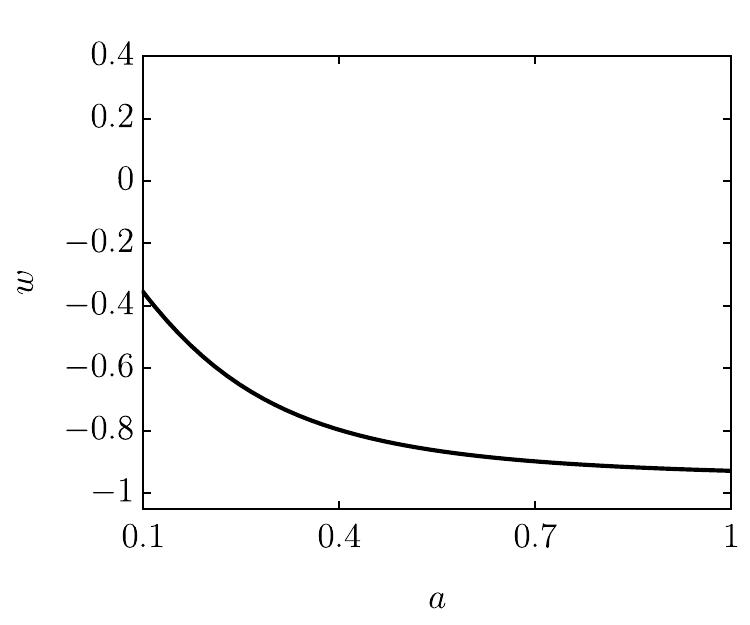}
\hspace*{-4.5cm}
\caption{(Left) Evolution of the equation of state $w(a)$. The time of transition to dark energy behavior is determined by $\hat{g}_Y$ and $\Omega_{EDE}$. (Right) Equation of state $w(a)$ from $a=0.1$ to present time, $a=1$. Although it resembles the behavior of a cosmological constant, the cosmic acceleration is only temporary.}
\label{fig:fig21}
\end{figure}

\begin{figure}[htp]
\centering
\hspace*{-5.5cm}
\includegraphics[width=0.4\linewidth]{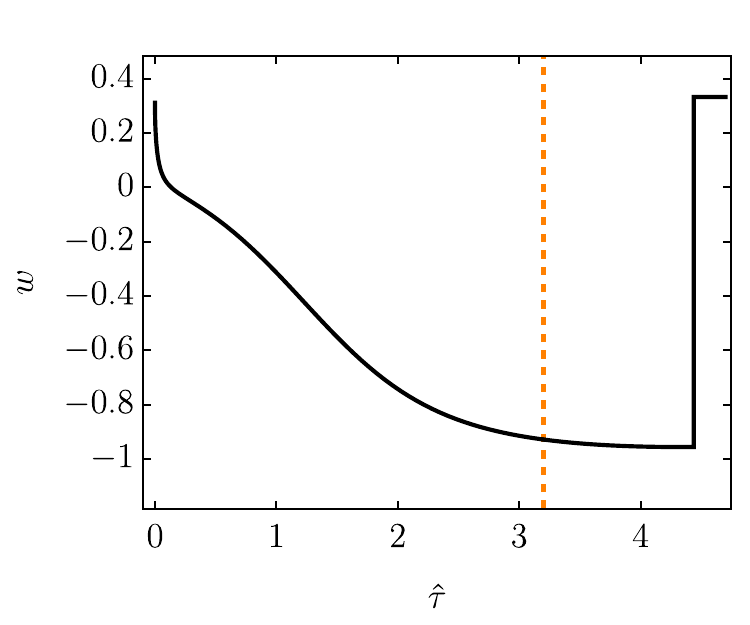}
\hspace*{1cm}
\includegraphics[width=0.425\linewidth]{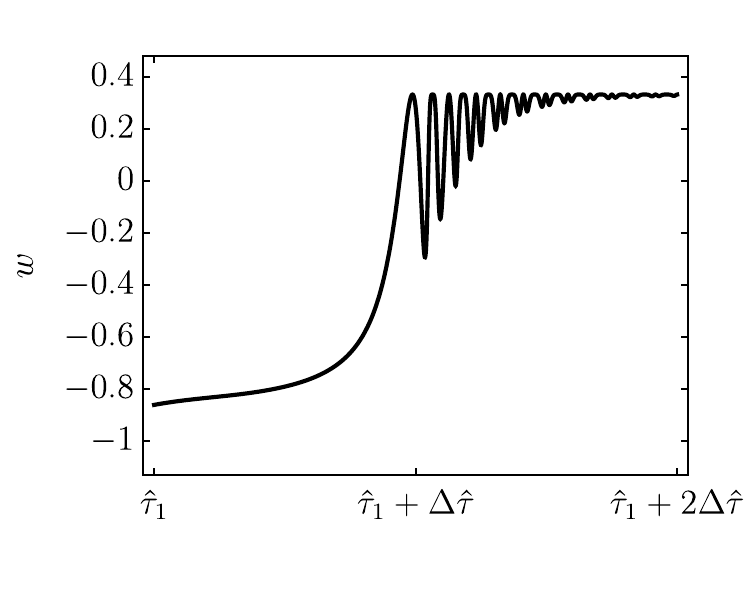}
\hspace*{-4.5cm}
\caption{(Left) Equation of state including future behavior, where it asymptotically increases back toward $1/3$. The vertical line at  $\hat \tau = 3.2$ indicates the present time. (Right) A more careful analysis reveals the oscillating behavior of $w$ as it becomes positive. Here $\hat \tau_1 = 4.5813$ and $\Delta \hat \tau = 2\times 10^{-6}$. In this example, the equation of state returns to $w=1/3$ in approximately $40$~Gyrs, at a future redshift $z\sim -0.78$.}
\label{fig:fig22}
\end{figure}

\begin{figure}[h]
    \includegraphics[width=0.475\linewidth]{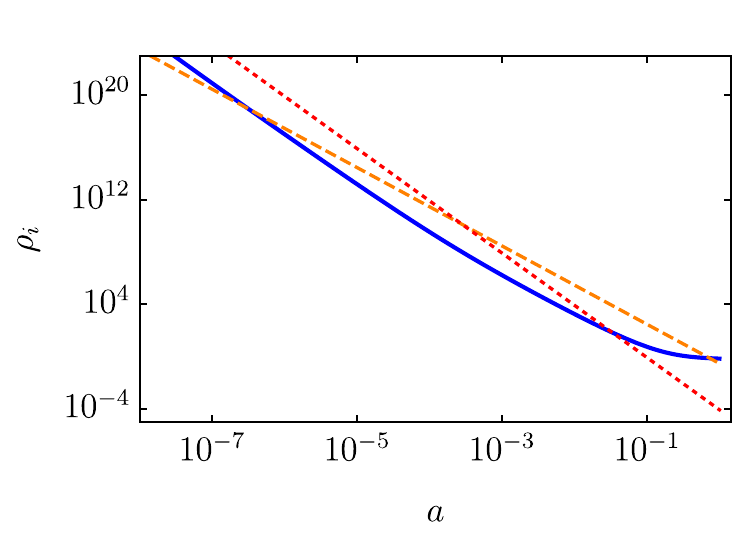}
    \includegraphics[width=0.45\linewidth]{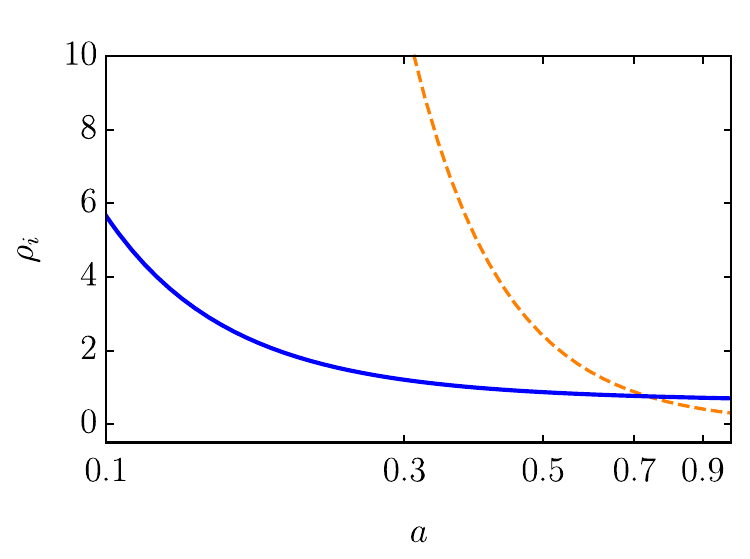}
\caption{(Left) Energy densities $\rho_i$ in units of the present-day critical energy density vs. scale factor $a$, showing $\rho_{D}$ (solid, blue), $\rho_R$ (dotted, red) and $\rho_M$ (dashed, orange). At early times the dark energy component tracks radiation until it reverses direction around $a=0.001$ and comes to dominate. (Right) Blow up of the cross over between matter and dark energy densities.}
\label{fig:fig23}
\end{figure}

The E solutions are obtained by setting $\theta \simeq \pi/2$. The equation of state starts at $w=1/3$, but it slowly evolves towards $-1$ as shown in Figs.~\ref{fig:fig21}-\ref{fig:fig22}, unlike for the B solution. In this case, the cosmic acceleration is temporary: the equation of state eventually returns to $1/3$ as shown in the right panel of Fig.~\ref{fig:fig22}, when the gauge field begins to oscillate. This return to radiation domination is in contrast with most freezing models of dark energy where the equation of state decreases toward -1, and only vaguely resembles most thawing models in which $w$ typically evolves towards matter domination \cite{Caldwell:2005tm}. Fig.~\ref{fig:fig23} shows the evolution of $\rho_{D}, \rho_R$ and $\rho_M$, the energy density of gauge quintessence, radiation, and matter, respectively. In general, increasing ${\hat g}_Y$ or $\Omega_{EDE}$ hastens the onset of oscillations of the gauge field, bringing the equation of state to $1/3$ sooner in the future. This behavior, linking the early and late regimes, plays a significant role in determining the observational constraints on this model.

\begin{figure}[h]
    \includegraphics[width=0.45\linewidth]{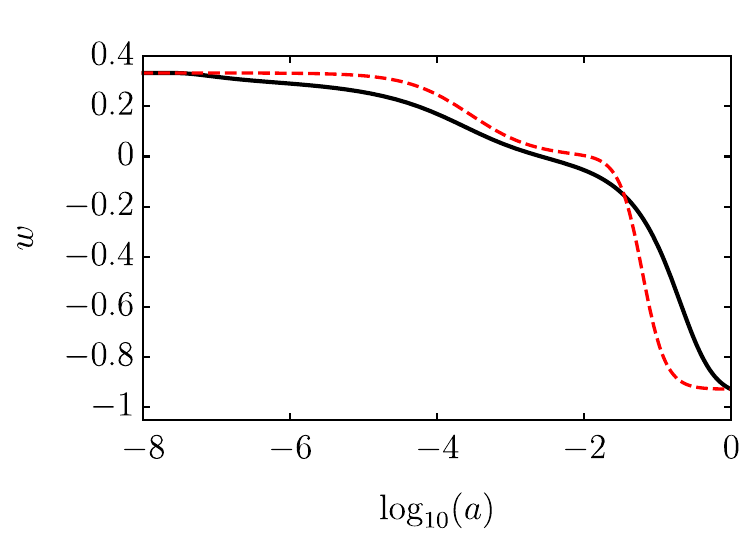}
    \includegraphics[width=0.45\linewidth]{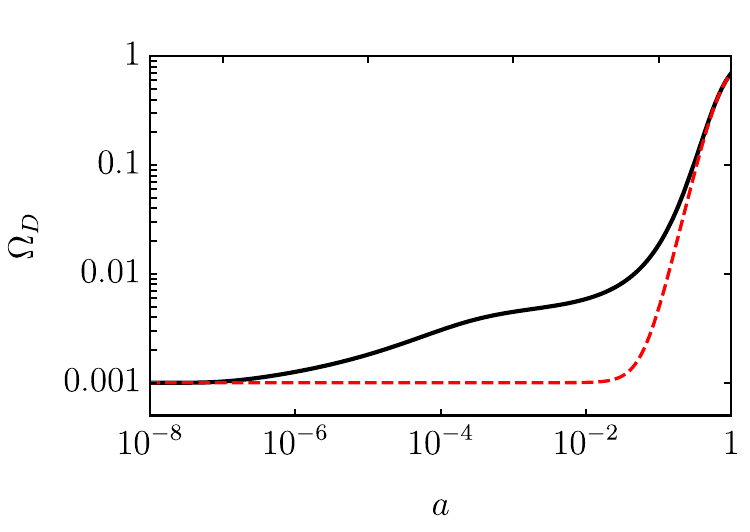}
\caption{(Left) Equation of state vs. a for the gauge quintessence model (black) and early dark energy analysis (dashed, red). (Right) Evolution of the relative abundance of dark energy for the gauge quintessence model (black) and early dark energy analysis (dashed, red). It can be seen that the relative abundance of dark energy grows more rapidly in our model.}
\label{fig:EDEmodel}
\end{figure}

It is interesting to compare the transition from radiation to dark energy behavior with a commonly studied EDE parametrization \cite{Doran:2006kp}. Therein, the fraction of dark energy at early times $\Omega_{EDE}$, its present day value $\Omega_D^0$ and $w_0$ as parameters are used to derive an analytical expression for the equation of state and abundance of dark energy as a function of $a$:
\begin{equation}
\Omega_D(a) = \frac{\Omega_D^0 - \Omega_{EDE}(1 - a^{-3 w_0})}{\Omega_D^0+ \Omega_M^0 a^{3 w_0}}+ \Omega_{EDE}(1 - a^{-3 w_0}).    
\end{equation}
In this parametrization, the dark component equation of state closely adheres to that of the background until late times. Comparison of the evolution of $w$ and $\Omega_D$ in our model with their predictions using $\Omega_{EDE} = 0.003$ is shown in Fig.~\ref{fig:EDEmodel}. In contrast, the equation of state of the gauge quintessence model only loosely adheres to that of the background. Moreover, the equation of state arrives near $w \sim -1$ very late, such that this model should be quite distinguishable from $\Lambda$CDM.

%%%%%%%%%%%%%%%%%%%%%%%%%%%%%%%%%%%%%%%%%%%%%%%%%%%%%%%%%%%
\subsection{Observational Constraints}

We now present constraints on the cosmological parameters of the E solutions of gauge quintessence using a combination of type 1a supernovae (SNe), baryon acoustic oscillations (BAO), and cosmic microwave background (CMB) data. These parameters consist of the present-day abundance, $\Omega_{DE} = 1 - \Omega_M$, the abundance deep in the radiation era $\Omega_{EDE}$ or alternatively as the abundance at recombination, $\Omega_{REC}$, and finally the Yang-Mills coupling, $\gy$. Our purpose is to determine a reasonable range of parameters which we can apply in our study of gravitational wave opacity.

Our constraint analysis proceeds from the following data sets. From SNe \cite{Betoule:2014frx} we constrain the luminosity distance - redshift relationship, $D_L(z)$. From BAO \cite{Aubourg:2014yra} we constrain the volume-averaged distance $D_V(z)$ and angular diameter distance $D_M(z)$, at several redshifts spanning $z \sim 0.1 - 0.6$, in units of the radius of the sound horizon at decoupling $r_d$. We use constraints to the baryon density, $\omega_b$, cold dark matter density, $\omega_c$, and the angular diameter distance to the horizon at decoupling from the CMB \cite{Hinshaw:2012aka,Ade:2013zuv}. This follows the same procedure as in Ref.~\cite{Aubourg:2014yra}. We also include the local determination of the Hubble constant \cite{Riess:2016jrr} which has contributed to the tension in the current data set \cite{DiValentino:2016hlg,Freedman:2017yms}. We recognize that this procedure is a gross simplification of the influence of gauge field dark energy on cosmological observables. In earlier work, however, we evaluated the effect of scalar and tensor perturbations of a cosmic gauge field as dark radiation on the CMB \cite{Bielefeld:2014nza,Bielefeld:2015daa}. There, we found that the dark radiation, present in abundances consistent with the $\Delta N_{\nu}$ bounds from Big Bang Nucleosynthesis and CMB recombination physics, did not leave a significant imprint on the CMB anisotropy pattern. We are also ignoring the imprint of the integrated Sachs-Wolfe (ISW) effect on the CMB, which arises from the change to the late-time evolution of the gravitational potentials at the onset of dark-energy domination. However, we note that the other constraints keep the late-time equation of state of the dark energy close to $-1$. This means the ISW under these models will not be distinctive relative to the case of $\Lambda$CDM. For the time being, we feel justified in this simple analysis, and we expect that parameters within the constraint boundaries would stand up to a more rigorous analysis.

\begin{figure}[htp]
\centering
\hspace*{-5.5cm}
\includegraphics[width=0.4\linewidth]{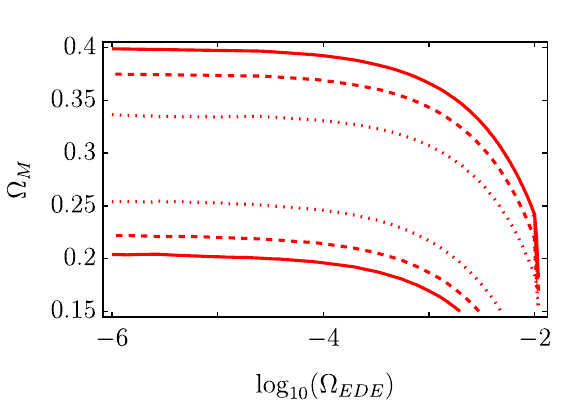}
\hspace*{0cm}
\includegraphics[width=0.4\linewidth]{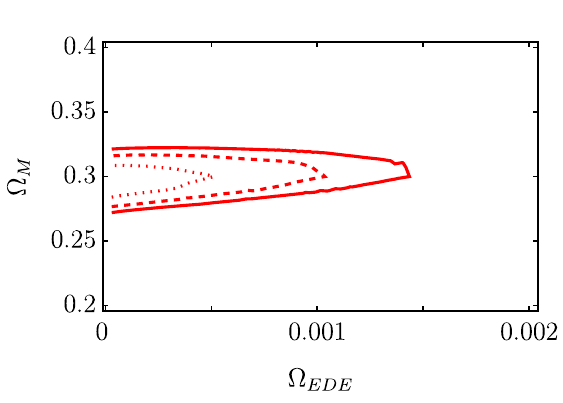}
\hspace*{-4.5cm}
\caption{$\Omega_{M}$ vs. $\Omega_{EDE}$ contours, for the case ${\hat g}_Y=1$ are shown due to SNe (left) and BAO and CMB (right). The dotted, dashed and solid lines represent the 1-, 2- and 3-$\sigma$ contours, respectively.}
\label{fig:fig3.1}
\end{figure}

\begin{figure}[htp]
\centering
\hspace*{-5.5cm}
\includegraphics[width=0.4\linewidth]{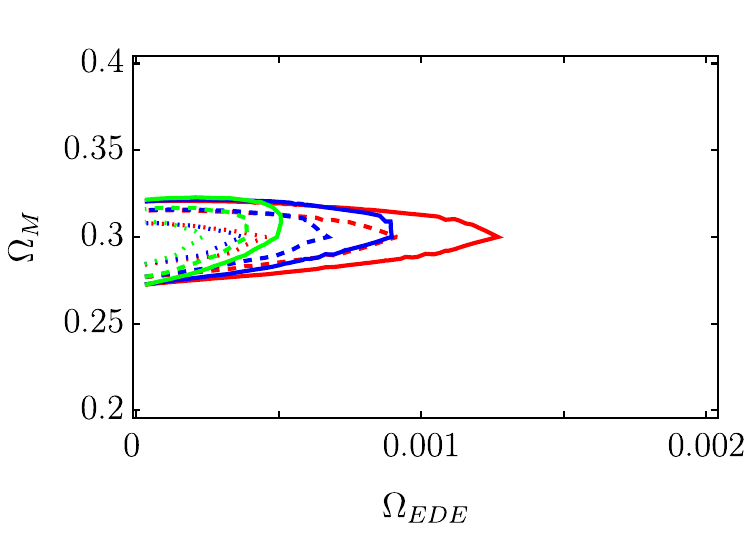}
\hspace*{0cm}
\includegraphics[width=0.4\linewidth]{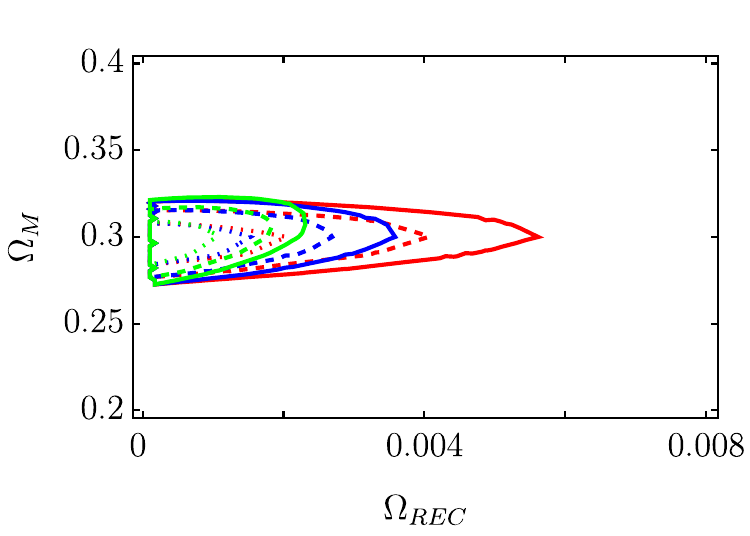}
\hspace*{-4.5cm}
\caption{The joint SNe, BAO, and CMB constraints on the $\Omega_M$ vs. $\Omega_{EDE}$ (left) or $\Omega_{REC}$ (right) parameter space for ${\hat g}_{Y} = 1,\, 5,\, 10$ in red, blue, and green.}
\label{fig:fig3.2}
\end{figure}

Contours for various combinations of these data sets are shown in Figs.~\ref{fig:fig3.1}-\ref{fig:fig3.2}. In Fig.~\ref{fig:fig3.1} we show the supernova luminosity distance - redshift constraint (left) separately from the BAO and CMB constraints (right). Here, we have set the Yang-Mills coupling to ${\hat g}_{Y} = 1$. For the SNe, the constraint depends only on $\Omega_M$, $\Omega_{EDE}$, and ${\hat g}_{Y}$. For the BAO and CMB, we have marginalized over parameters $H_0,\, \omega_b,\,\omega_c$. The dotted, dashed and solid lines represent the 1-, 2- and 3-$\sigma$ contours, respectively. The BAO and CMB clearly give a tighter constraint, but the SNe are important for marking the boundary at the rightmost tip of the constraint region when the two curves are combined. Fig.~\ref{fig:fig3.2} shows the combined constraints for three values of the Yang-Mills coupling, ${\hat g}_{Y} = 1,\, 5,\, 10$ in red, blue, and green. The two panels show the same constraint regions with the horizontal axis marking either $\Omega_{EDE}$ (left) or $\Omega_{REC}$ (right). We note that increasing ${\hat g}_Y$ shrinks the constraint region; stronger coupling shortens the duration of the accelerating epoch, but may be compensated for by decreasing the early-time abundance. Further decreasing ${\hat g}_Y < 1$ has the effect of continuing the accelerating era into the future, but otherwise has no effect on the constraint region.

The results shown here for an SU(2) gauge field can be generalized to SU(N) with ${\cal N}$ disjoint subgroups, as described in the Appendix. In short, the results for ${\cal N}$ subgroups, each with ${\hat g}_{Y{\cal N}}$, are equivalent to the results for a single SU(2) with ${\hat g}_{Y1}={\hat g}_{Y{\cal N}}/\sqrt{\cal N}$. Hence, increasing the number of fields slightly extends the parameter region to the right, with diminishing effect as ${\hat g}_{Y1}$ approaches unity from above.

%%%%%%%%%%%%%%%%%%%%%%%%%%%%%%%%%%%%%%%%%%%%%%%%%%%%%%%%%%%
\section{Modulation of Gravitational Wave Amplitude}
\label{sec4}

We now return to the phenomenon of GWGF oscillations and investigate it in the context of the E and B solutions of the gauge quintessence model as described by the action Eq.~(\ref{eqn:action}). We start with a presentation of the full gravitational and gauge field wave equations of motion. We again consider $z-$directed gravitational and gauge field waves as given in Eqs.~(\ref{eqn:Atnsr1}-\ref{eqn:Atnsr2}). To put the action in canonical form, we write
\begin{equation}
h_{+,\times} = \frac{\sqrt{2}}{a M_P}v_{+,\times}, \qquad y_{+,\times} = \frac{1}{\sqrt{2}a}u_{+,\times}
\end{equation}
and furthermore to decouple the equations, we switch to the circular polarization basis
\begin{eqnarray}
v_{+} &=& \frac{1}{\sqrt{2}}(v_L + v_R), \qquad v_{\times} = \frac{i}{\sqrt{2}}(v_L-v_R), \cr
u_{+} &=& \frac{1}{\sqrt{2}}(u_L + u_R), \qquad u_{\times} = \frac{i}{\sqrt{2}}(u_L-u_R).
\end{eqnarray}
The full equations, in terms of the Fourier space amplitudes, are as follows:
\begin{eqnarray}
&&v_R'' + \left[k^2 - \frac{a''}{a}  + \frac{2}{a^2 M_P^2}(\gy^2 \phi^4 - \phi'^2)\right] v_R = \frac{2}{a M_P}\left[ (\gy \phi - k)\gy  \phi^2 u_R -  \phi' u_R'\right]
\label{eqn:vR} \\
&&u_R'' + \left[k^2 - 2 \gy k \phi - \frac{\kappa }{g  M_P^4}(\gy \phi - k)\left(\frac{\phi^2 \phi'}{a^4}\right)'\right]u_R  
=\frac{2}{a M_P}\left[ a\left(\frac{\phi' }{a}v_R\right)' + (\gy\phi-k) \gy\phi^2 v_R  \right].
\label{eqn:uR}
\end{eqnarray}
The equations for $v_L,\,u_L$ are obtained by replacing $k \to -k$ or $\gy \to -\gy$.

In Appendix \ref{app1}, we generalize these equations to the case where the action features ${\cal N}$ disjoint SU(2) subgroups. In this case the energy density and pressure become
\begin{equation}
\rho = \frac{3}{2}{\cal N}\left(\frac{\phi'^2}{ a^4} + \gy^2 \frac{\phi^4}{ a^4}\right) + \frac{3}{2}{\cal N}^2\frac{\kappa}{M_P^4} \left(  \frac{\phi' \phi^2}{a^4}\right)^2, \qquad
p = \frac{1}{2}{\cal N}\left(\frac{\phi'^2}{ a^4} + \gy^2 \frac{\phi^4}{ a^4}\right) -  \frac{3}{2}{\cal N}^2\frac{\kappa}{M_P^4} \left(  \frac{\phi' \phi^2}{a^4}\right)^2 
\end{equation}
and the equation of motion is
\begin{equation}
\phi'' + 2 \gy^2 \phi^3 + \frac{\kappa}{M_P^4 a^4} {\cal N} (\phi^4 \phi'' + 2 \phi^3 \phi'^2 -4 (a'/a)\phi^4 \phi')=0.
\end{equation}
The gravitational and gauge field wave equations become
\begin{eqnarray}
&&v_R'' + \left[k^2 - \frac{a''}{a}  + \frac{2 {\cal N}}{a^2 M_P^2}(\gy^2 \phi^4 - \phi'^2)\right] v_R = \sum_{n=1}^{\cal N}\frac{2}{a M_P}\left[ (\gy \phi - k)\gy  \phi^2 u_{Rn} -  \phi' u_{Rn}'\right]
\label{eqn:vRn} \\
&&u_{Rn}'' + \left[k^2 - 2 \gy k \phi - \frac{\kappa }{\gy  M_P^4}{\cal N}(\gy \phi - k)\left(\frac{\phi^2 \phi'}{a^4}\right)'\right]u_{Rn}  
=\frac{2}{a M_P}\left[ a\left(\frac{\phi' }{a}v_R\right)' + (\gy\phi-k) \gy \phi^2 v_R  \right].
\label{eqn:uRn}
\end{eqnarray}
As derived in Ref.~\cite{Caldwell:2016sut}, we obtain the equations of motion in the high-frequency limit by making the ansatz $v = {\cal h}e^{-i k \tau}$ and $u_n = {\cal y}_n e^{-i k \tau}$, whereupon the amplitudes evolve as
\begin{equation}
{\cal h}'_R = -\frac{1}{a M_P}\sum_n(\phi' + i \gy \phi^2) {\cal y}_{Rn}, \qquad
{\cal y}'_{Rn} = \frac{1}{a M_P}(\phi' - i \gy \phi^2) {\cal h}_R + i \gy \phi {\cal y}_{Rn} - 
\frac{i \kappa {\cal N} }{2 \gy  M_P^4}\left(\phi^2 \phi'/a^4\right)' {\cal y}_{Rn}.
\label{eqn:hifreq}
\end{equation}
The equations for the left-handed polarization are obtained by taking $\gy \to -\gy$, which is equivalent to taking the complex conjugate of the equations of motion. Hence, the results in terms of the wave amplitude are the same for either chirality.

The background equations for the case ${\cal N}>1$ are equivalent to the SU(2) case when the following identifications are made:
\begin{equation}
\phi_{\cal N} = \frac{1}{\sqrt{\cal N}}\phi_1, \qquad g_{\cal N} = \sqrt{\cal N} g_{1}, \qquad \kappa_{\cal N} = {\cal N} \kappa_{1}.
\end{equation}
Furthermore, when all the gauge field waves behave identically, such that ${\cal y}_1 = {\cal y}_2...$, then we define ${\cal y}_{\cal N} \equiv {\cal N}^{-1}\sum {\cal y}_n$, such that if we rescale
\begin{equation}
{\cal y}_{\cal N} = \frac{1}{\sqrt{\cal N}} {\cal y}_{1}
\end{equation}
then the high frequency equations of motion are equivalent to the SU(2) case.

We solve the system of equations (\ref{eqn:hifreq}) describing high frequency gravitational and tensor gauge field waves. We use boundary conditions ${\cal h}=1$ and ${\cal y}=0$ to determine the gravitational wave opacity due to the gauge field. We can also see the origin of the coefficients referred to earlier as $b_1 = \phi'/a M_P$, $b_2 = \gy \phi^2/a M_P$, and $b_3 = \gy \phi$. However, there is a new term in Eq.~(\ref{eqn:hifreq}) that was not present in the simplified model of Sec~\ref{sec2}, namely the $\kappa$-dependent term. For consistency of our language, we redefine 
\begin{equation}
b_3 = \gy \phi - \frac{\kappa {\cal N}}{\gy M_P^4}\left( \phi^2 \phi'/a^4\right)'.
\end{equation}
Inspecting Eq.~(\ref{eqn:hifreq}), we determine that these $b_i$ coefficients must be non-negligible relative to $a'/a$,  and $b_3$ not so much larger than $b_1,\,b_2$, in order for the gauge field wave to grow and thereby modulate the gravitational wave amplitude.

%%%%%%%%%%%%%%%%%%%%%%%%%%%%%%%%%%%%%%%%%%%%%%%%%%%%%%%%%%%
\subsection{Gravitational Wave Opacity under B Solutions}
\label{sec4b}

\begin{figure}[htb]
\centering
\includegraphics[width=0.45\linewidth]{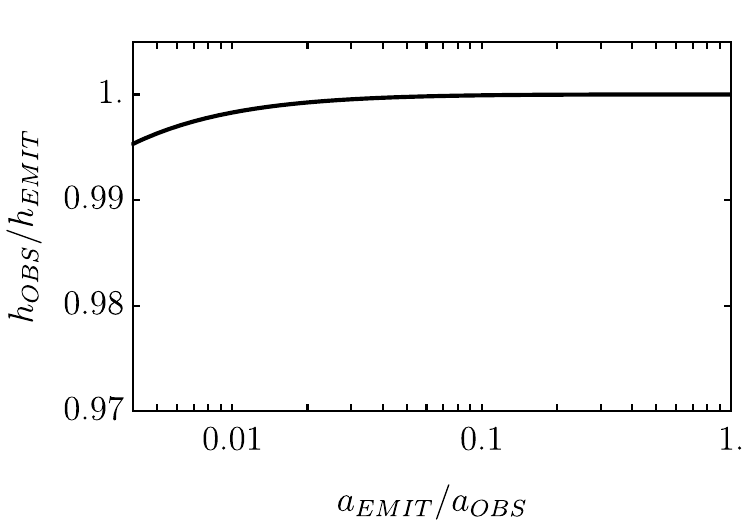}
\caption{The high frequency gravitational wave transmission amplitude is shown for the case of a sample B solution, also illustrated in Fig.~\ref{fig:bsoln}. In this case, as with all B solutions, the effect of the gauge field is negligible and the transmission is nearly $100\%$. Only for GWs that originate at $z \gtrsim 250$ is there $\sim 0.5\%$ suppression.}
\label{fig:opacityb}
\end{figure}

The high-frequency gravitational wave transmission is nearly $100\%$, meaning the opacity is nearly zero, in the presence of the B solutions of the gauge quintessence equations of motion. An example is shown in Fig.~\ref{fig:opacityb}. The reason for the weak effect is that the dark energy equation of state $w$ is so close to $-1$ that it suppresses the radiation portion of the gauge field energy density, with $b_1,\,b_2 \ll b_3$. Consequently, the gauge field wave $y$ is unexcited, thereby leaving the gravitational waves free to propagate. 
 
%%%%%%%%%%%%%%%%%%%%%%%%%%%%%%%%%%%%%%%%%%%%%%%%%%%%%%%%%%%
\subsection{Gravitational Wave Opacity under E Solutions}
\label{sec4e}

\begin{figure}[htb]
\centering
\includegraphics[width=0.45\linewidth]{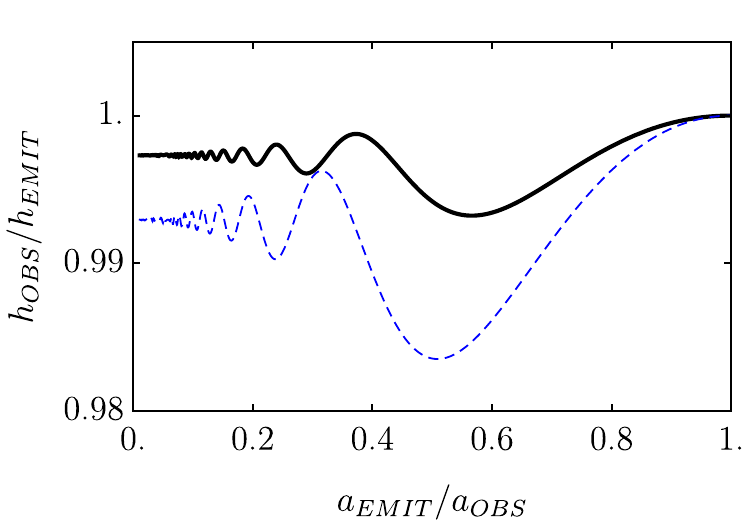}
\caption{The high frequency gravitational wave transmission amplitude is shown for the case of a sample E solution. The solid (black) curve is for a model with $\Omega_M=0.3$, ${\hat g}_{Y}=1$ and $\Omega_{EDE}=0.0013$, lying on the boundary of the $2\sigma$ contour, illustrated in Fig.~\ref{fig:fig3.2}. The effect is small, with less than $1\%$ dimming of standard sirens emitted near $z=1$. For comparison, the dashed (blue) curve shows the transmission for a model with $\Omega_M=0.3$, ${\hat g}_{Y}=1$, and $\Omega_{EDE}=0.003$, which is incompatible with the observational constraints.}
\label{fig:opacitye}
\end{figure}

We naively expect the gravitational wave transmission to decrease by $\sim 1-10\%$ in the presence of E solutions, since the gauge field energy density is non-negligible until late times. It is easy to find numerical examples of models that are superficially consistent with data -- e.g. negative equation of state at present, abundant in the radiation era, consistent with the bound on $\Delta N_\nu$ -- which lead to a significant suppression of gravitational wave amplitude. However, a careful analysis of observational constraints leads to a different story. The strongest possible effect in this scenario is obtained when $\Omega_{EDE}$ is largest. The high-frequency gravitational wave transmission dips by less than $\sim 1\%$ in the presence of the E solutions of the gauge quintessence equations of motion that are also compatible with the observational constraints. The effect, illustrated in Fig.~\ref{fig:opacitye}, is peaked at $z\simeq 1$. In the absence of this resonant conversion of gravitational waves into gauge field waves, the amplitude should be constant since we have already accounted for the redshifting of the wave amplitude. For comparison, we also show the opacity for a model with $\Omega_{EDE}$ that exceeds the observational bounds. The effect in the case of the E solutions is greater than for the B solutions because the coefficient $b_3$ is not so much larger than $b_1,\,b_2$; that is, the radiation portion of the gauge field energy density is non-negligible.  However, because the gauge field equation of state only slowly interpolates between $1/3$ and $-1$ in the matter era, the effect on the expansion history pushes up against the observational constraints. Hence, the tight constraint on $\Omega_{EDE}$ keeps compatibility with the angular-diameter distance to the CMB and also suppresses the effect on gravitational waves.

%%%%%%%%%%%%%%%%%%%%%%%%%%%%%%%%%%%%%%%%%%%%%%%%%%%%%%%%%%%
\subsection{Effect on Long Wavelength Gravitational Waves}
\label{sec4g}

The gauge field also has an effect on long wavelength gravitational waves. This has previously been explored in Refs.~\cite{Bielefeld:2014nza,Bielefeld:2015daa} wherein the gauge field contributes dark radiation. The case of gauge field dark energy as studied in this paper is new, although the results for long wavelengths are similar, as we now describe. First, we note that Eq.~(\ref{eqn:uRn}) includes an effective mass term that depends on the difference, $B^2-E^2$. This does not affect the speed of propagation, but it does lead to amplification (suppression) of any long wavelength primordial gravitational waves for the E (B) solutions. The effect depends on the relative abundance of the gauge field dark energy, particularly during the radiation-dominated era. Due to the constraints on additional relativistic species, the amplification (suppression) in time would be manifest as a slight red (blue) tilt in wavenumber. Second, returning to Eqs.~(\ref{eqn:vRn}-\ref{eqn:uRn}), we see that the coupling between the gauge field and gravitational waves remains even for long wavelengths. However, the gauge field ``stiffens" on long wavelengths as the equation of state approaches $w\to -1$. To see this, note that the quantity $(\phi^2 \phi' / a^4)$ approaches a constant when $w\to -1$, since this is the exact combination that is responsible for the dark energy. This means the effect on long wavelengths will diminish with the onset of cosmic acceleration, and we can refer to the dark radiation scenario for guidance as to the behavior. We also expect that the effect on scalar and vector perturbations will diminish as the gauge field stiffens; since this occurs before the dark energy becomes dominant, gauge field inhomogeneities are not expected to play a significant role in the large scale structure and CMB anisotropy, as we mentioned earlier. Since our current focus is the gravitational wave opacity effect, we will consider the inhomogeneities in future work.

%%%%%%%%%%%%%%%%%%%%%%%%%%%%%%%%%%%%%%%%%%%%%%%%%%%%%%%%%%%
\subsection{Detection}
\label{sec4f}
 
Will future gravitational wave observatories be able to detect the $\sim1\%$ modulation that we predict in this dark energy scenario? There are many new GW observatories planned for the next decades, but only a few with the sensitivity and frequency range to meet our needs. This effect is strongest for GWs emitted at a scale factor $a/a_0\sim 0.5$ or $z \sim 1$, as seen in Fig.~\ref{fig:opacitye}. Hence, detection requires a well-measured GW standard siren and a well-measured source redshift in the vicinity of $z\sim 1$.

The most reliable GW sources foreseen as standard sirens are binary neutron stars (BNS) such as already detected by LIGO/Virgo \cite{TheLIGOScientific:2017qsa}, as well as massive black hole binaries (MBHB) that will comprise one of the primary targets of LISA. The Hubble diagram using BNS standard sirens is expected to come into sharp focus over the next decade, as the LIGO/Virgo network is extended to include sites in India and Japan, and new detector technologies are implemented \cite{Reitze:2019dyk}. However, the reach of these observatories is only expected to be $z \sim 0.2$ and will not be sufficient to test our theoretical models \cite{Chen:2017wpg}. Instead, we look ahead to the network of proposed third-generation detectors (3G) such as the Cosmic Explorer (CE) in the US \cite{Evans:2016mbw,Reitze:2019iox} or the Einstein Telescope (ET) in Europe \cite{Sathyaprakash:2009xt,Punturo:2010zz}.

Significantly advanced 3G ground-based experiments have been proposed for the 2030s. 3G experiments are forecast to have a horizon $z > 10$ and a response distance $d_{50}$ corresponding to $z=1.9$, meaning that half the objects at this distance, allowing for random orientations on the sky, will be detected \cite{Chen:2017wpg}. Moreover it is forecast that a network of 3G observatories such as CE and ET will observe many BNS sirens with sufficient angular resolution to localize the source to within a reasonably narrow window on the sky, adequate for EM follow-up. We consider a heterogeneous CE-ET network consisting of two L-shaped, 40~km CE detectors, one in the US and one in Australia, and one 10~km, triangular ET observatory in Europe. Mills {\it et al} \cite{Mills:2017urp} estimate that such a network can detect $1.3 \times 10^6$ BNS mergers, with $3 \times 10^5$ localized within 10 square degrees, per year. We use the model presented in Ref.~\cite{Mills:2017urp} for the number of BNS observations and localizations per year with future networks, as a function of redshift and luminosity distance. For the sake of argument, we consider all $3 \times 10^5$ BNS coalescences that are localized within 10 square degrees to have well measured EM redshifts. We assume all mergers are resolved at better than SNR$=10$. We consider that the most significant source of error for this population is due to weak gravitational lensing, using the model of lensing distortion presented in \cite{Hirata:2010ba}. To assess the detectability of the GW opacity, we evaluate the difference in standard siren luminosity distances with and without the GW opacity: $\Delta d_L(z) = d_L({\rm with}) - d_L({\rm w/o})$. We thereby compute
\begin{equation}
{\rm SNR}^2 = \int dz \, \frac{dN}{dz} \, \frac{ \Delta d_L^2 }{ \sigma_{d_L}^2 }
\end{equation}
where $\sigma_{d_L}^2$ is given in  \cite{Tamanini:2016zlh}, and we read $dN/dz$ off Figure 5 of Ref.~\cite{Mills:2017urp}. Using this simple model, we estimate ${\rm SNR} \simeq 15,\, 35$ for the two models illustrated in Fig.~\ref{fig:opacitye}. Hence, under the idealized circumstances represented by this model, the novel effect is distinguishable. We will carry out a more detailed study elsewhere.

LISA, expected to launch in the mid-2030s, will extend the horizon for standard sirens to $z\sim 10$ or greater, using MBHBs \cite{Audley:2017drz}. These are massive black holes, ranging from $10^4 - 10^7\, M_\odot$, which are expected to merge in accretion-gas rich environments, thereby producing an EM counterpart. However, the number of systems for which an EM redshift can be obtained is forecast to be less than 100 in a 4 year mission \cite{Tamanini:2016zlh}. While very promising for the range of redshift, this method would appear to be inadequate to accumulate sufficient signal to distinguish the effect of GW opacity in our theoretical models.

We have not considered the use of so-called dark standard sirens, due to binary black hole (BBH) mergers that have no EM counterpart to provide a redshift. These signals can still be used for cosmography either by probabilistically identifying a well-localized source with a host galaxy in a galaxy redshift catalog \cite{Soares-Santos:2019irc} or by using some prior knowledge about the source population redshift distribution \cite{Farr:2019twy}. On an individual basis, BBH standard sirens are not as precise as systems with an EM counterpart, and so, for the time being, we have concentrated on BNS and MBHB sirens.

%%%%%%%%%%%%%%%%%%%%%%%%%%%%%%%%%%%%%%%%%%%%%%%%%%%%%%%%%%%
\section{Discussion}
\label{sec5}

In this work we have investigated a model of gauge field dark energy in which GWs are absorbed and re-emitted by the gauge field. This resonant conversion of energy translates into a distinct modulation pattern on the GW amplitude, which we describe as a redshift dependent opacity. Such an effect could have implications for the use of standard sirens to chart the cosmic expansion history. In the event the GW source can be coupled with an electromagnetic counterpart, it is expected that its luminosity distance can be inferred within $1-10\%$ uncertainty \cite{Hughes:2001ya}, which in turn can be used to constrain $H_0$ at the level of $0.5\%$ in the most favorable scenarios \cite{Zhao:2010sz,Cai:2016sby,Tamanini:2016zlh}. 

The dark energy model under consideration, gauge quintessence, features an equation of state $w=1/3$ at early times, but which evolves to $w \simeq -1$ by the present day. We have shown that there are two broad classes of gauge field evolution that realize these features, labelled B- and E-solutions. Whereas the equation of state under the B solution evolves sharply from $w=1/3$ to $w=-1$, the evolution under the E solution is softer. The consequences are twofold. First, the field evolution in the late universe under the E solution is greater than the under the B solution. This means that the rate of gravitational wave - gauge field oscillation is greater under E than B. Second, the softer evolution of $w$ under the E solutions means the gauge field energy density during the matter era is greater than under B. By evaluating the observational constraints based on SNe, CMB and BAO data, we have been able to determine the viable range of dark energy parameters. We used these results to study the opacity to weak gravitational waves in these models. 

An open question is whether the strong, time-varying gravitational fields that arise in black hole mergers can excite the gauge field dark energy. From an energetics point of view, the dark energy field carries far less energy in the volumes swept out by the merging sources than the energy of the merging stars themselves. For this reason, we suspect dark energy will have a very small effect on the waveforms created at merger. In contrast, by comparison, the GW-GF conversion effect that we model needs to accumulate over cosmological distances and time scales to produce a discernible effect.

Despite simple expectations that the opacity could have a $\sim 1-10\%$ effect on gravitational wave amplitudes, we find that for realistic models the dimming is smaller. For B solutions, the dimming is negligible. For E solutions, the dimming is up to $1\%$ for sources emitted near redshift $z\sim 1$. We have shown, based on a simplistic model of the BNS standard sirens, that the GW opacity effect may be detectable by a network of 3G observatories. Future work will follow up on the theoretical dark energy model and observational forecasts.

%%%%%%%%%%%%%%%%%%%%%%%%%%%%%%%%%%%%%%%%%%%%%%%%%%%%%%%%%%%%
\acknowledgments
We thank Eric Linder and an anonymous referee for useful comments. This work is supported in part by DOE grant DE-SC0010386. 

%%%%%%%%%%%%%%%%%%%%%%%%%%%%%%%%%%%%%%%%%%%%%%%%%%%%%%%%%%%%

%%%%%%%%%%%%%%%%%%%%%%%%%%%%%%%%%%%%%%%%%%%%%%%%%%%%%%%%%%%
\appendix

%%%%%%%%%%%%%%%%%%%%%%%%%%%%%%%%%%%%%%%%%%%%%%%%%%%%%%%%%%%
\section{Multiple Gauge Field Quintessence}
\label{app1}

By taking advantage of the fact that multiple SU(2) subgroups may be embedded in a larger SU(N) group, the gauge quintessence model can be generalized to include an arbitrary number of gauge-fields. In this section, we give an overview of the mathematical procedure behind this approach, effectively introducing an extra degree of freedom $\cal{N}$ corresponding to the number of fields in the model.

In the single field case, the requirement of Eqn.~(\ref{eqn:fsl}) ensures that the flavor components of $A_\mu$ match the generators of the SU(2) group that make up the canonical basis of its algebra -- namely, the Pauli matrices. Specifically, each Pauli matrix $\sigma_i$ is uniquely associated with one of the vectors $A_\mu$, where $\mu \neq 0$, so that $A^i_i = \phi(t)$. Note that the choice $\mu = i$ is not essential to guarantee isotropy as the resulting equations of motion remain unchanged in all six possible alignments.

In fact, rewriting equation Eqn.~(\ref{eqn:fdefn}),
\begin{equation}
F^a_{\mu \nu} = \partial_\mu A^a_\nu - \partial_\nu A^a_\mu - \gy f^a_{bc} A^b_\mu A^c_\nu
\end{equation}
we note that the theory remains unchanged for different SU(N) gauges, provided $A^a_\mu$ is constructed according to one of the $N^2 -1$ dimensional bases of $\mathfrak{su}(2)$ -- the algebra of SU(2). Thus, while $a$ now runs from $1, \ldots, N^2 -1$, only those components of $A^\mu$ which are mapped to an SU(2) subgroup of SU(N) will contribute to the field strength, and will do so identically to the $N=2$ case.

The larger dimensionality, however, gives more freedom in choosing a particular basis of generators, since we are no longer limited to the Pauli matrices but may instead pick another basis to populate our gauge field. For example, with SU(3), one may define $A^a_\mu$ as either one of the following two matrices (omitting the time component):
\begin{equation}
A_i^a= \phi(t) \left( \begin{array}{ccc}
1 & 0 & 0  \\
0 & 1 &0 \\
0 & 0 & 1 \\
0 & 0 & 0 \\ 
0 & 0 & 0 \\ 
0 & 0 & 0  \\
0 & 0 & 0  \\
0 & 0 & 0  \\
\end{array} \right),
\, \, \,
A_i^a= \phi(t) \left( \begin{array}{ccc}
0 & 0 & 0  \\
0 & 0 &0 \\
0 & 0 & -\frac{1}{2}  \\
0 & 0 & 0 \\ 
0 & 0 & 0 \\ 
1 & 0 & 0  \\
0 & 1 & 0  \\
0 & 0 & \frac{\sqrt{3}}{2}  \\
\end{array} \right)
\end{equation}
The second arrangement corresponds to the following basis
\begin{equation}
\{ \lambda_6, \lambda_7, \frac{1}{2} (-\lambda_3+ \sqrt{3} \lambda_8) \}
\end{equation}
where the $\lambda_i$ denote the Gell-Mann matrices. One will also recognize it as the basis used in constructing one of the standard SU(3) ladder operators:
\begin{align}
U_\pm = \frac{1}{2} ( \lambda_6 \pm i \lambda_7 ), \, \,  U_z = \frac{1}{2} (  - \lambda_3 + \sqrt{3} \lambda_8)
\end{align}

Since a basis for $\mathfrak{su}(3)$ can be defined so as to carry over the Pauli matrices in the form of $2 \times 2$ block matrices embedded in $3 \times 3$ generators, we effectively end up with two SU(2) subgroups to choose from. Though more combinations of generators obeying the SU(2) commutation relations can be found, only those corresponding to the $N-1$ basis elements of the Cartan subalgebra will here be considered, and henceforth referred to as independent.

In order to include more than one field in our action, the gauge field is taken as the sum of two or more subcomponents, each populated according to its own corresponding SU(2) subgroup:
\begin{equation}
A^a_\mu = A^a_\mu(1) + A^a_\mu(2) + A^a_\mu(3) + \ldots
\end{equation}
where each individual $A^a_\mu(i)$ is labeled according to which independent SU(2) subgroup  it corresponds to. In SU(4), there are 3 such subgroups, namely
\begin{align}
S_1 &= \{ \lambda_1, \lambda_2, \lambda_3 \} \\
S_2 &= \{ \lambda_6, \lambda_7, \frac{1}{2} (-\lambda_3+ \sqrt{3} \lambda_8) \} \\
S_3 &= \{ \lambda_{13}, \lambda_{14}, \frac{1}{3} (-\sqrt{3} \lambda_8+ \sqrt{6} \lambda_{15}) \}
\end{align}
where each generator making up the $15$-dimensional basis of  $\mathfrak{su}(4)$ is now a $4 \times 4$ matrix. 

Furthermore, in building $A^a_\mu$, only those $A^a_\mu(i)$ whose corresponding bases $S_i$ are disjoint may be combined, or else isotropy will be spoiled in the direction of overlap. For example, an SU(3) gauge-field built out of $ A^a_\mu(1)$ and $ A^a_\mu(2)$ will result in an anisotropic stress-energy tensor as the pressure component in the $z$ direction will differ from the other two -- a direct consequence of the fact that $S_1 \cap S_2 = \lambda_3$. Hence, for a two-field configuration, we consider the SU(4) gauge, where
\begin{equation}
A^a_\mu = A^a_\mu(1) + A^a_\mu(3)
\end{equation}
This puts a further restriction on the maximum number of independent, non-overlapping gauge fields $\cal{N}$ one can form out for each of the $N-1$ Cartan elements in SU(N), and it can easily be shown that $\cal{N}$ is the greatest integer such that $\cal{N} \leq$ $N/2$. Therefore, our model can accommodate up to two fields in an SU(4) and SU(5) gauge, three in an SU(6) and SU(7) gauge, etc. 

This procedure can be generalized to any SU(N) group, and thus any number of field $\cal{N}$. In order to devise a basis for $\mathfrak{su}(N)$, we seek $N^2 - 1$ linearly independent traceless Hermitian matrices. If we let $M_{mn}$ denote the $N \times N$ matrix with 1 in the $m^{\text{th}}$ row and $n^{\text{th}}$ column and zeroes elsewhere, these matrices may be constructed as follows:
\begin{equation}
\begin{aligned}
&M_{nm} + M_{mn} && (1 \leq m < n \leq N)  \\
&i M_{nm} - i M_{mn} &&  ( 1 \leq m < n \leq N) \\
& (I - N M_{nn})/A_n && (n \leq N)
\end{aligned}
\label{eqn:basis}
\end{equation}
where $I$ is the $N \times N$ identity matrix and
\begin{equation}
A_N \equiv \sqrt{ \frac{(N-1) + (N-1)^2}{2} }
\end{equation}
is the normalization factor. The extra SU(2) subgroup can then be constructed similarly by considering the ladder operator corresponding to the extra Cartan element given by $ (I - N M_{NN})/\sqrt{A_N}$, and which in general can be associated with the following basis:
\begin{equation}
S_{N-1} = \Big\{ \lambda_{N^2 -3}, \lambda_{N^2 -2}, \frac{1}{N-1} \left( A_N \lambda_{N^2 -1} - A_{N-1} \lambda_{(N-1)^2 -1 } \right) \Big\}
\end{equation}
where generators are labelled according to the ordering prescribed by Eqn.~(\ref{eqn:basis}).

In the case of ${\cal N}$ disjoint SU(2) subgroups, of which ${\cal N}_R$ are right handed and ${\cal N}_L$ are left handed, the energy density and pressure become
\begin{equation}
\rho = \frac{3}{2}{\cal N}\left(\frac{\phi'^2}{ a^4} + \gy^2 \frac{\phi^4}{ a^4}\right) + \frac{3}{2}\Delta{\cal N}^2\frac{\kappa}{M_P^4} \left(  \frac{\phi' \phi^2}{a^4}\right)^2, \qquad
p = \frac{1}{2}{\cal N}\left(\frac{\phi'^2}{ a^4} + \gy^2 \frac{\phi^4}{ a^4}\right) -  \frac{3}{2}\Delta{\cal N}^2\frac{\kappa}{M_P^4} \left(  \frac{\phi' \phi^2}{a^4}\right)^2 
\end{equation}
where $\Delta{\cal N} = {\cal N}_R - {\cal N}_L$. The equation of motion is
\begin{equation}
\phi'' + 2 \gy^2 \phi^3 + \frac{\kappa}{a^4 M_P^4}\frac{\Delta{\cal N}^2}{{\cal N}} (\phi^4 \phi'' + 2 \phi^3 \phi'^2 -4 (a'/a)\phi^4 \phi')=0.
\end{equation}
The gravitational wave equation becomes
\begin{eqnarray}
v_R'' + \left[k^2 - \frac{a''}{a}  + \frac{2 {\cal N}}{a^2 M_P^2}(\gy^2 \phi^4 - \phi'^2)\right] v_R &=& \sum_{n=1}^{{\cal N}_R}\frac{2}{a M_P}\left[ (\gy \phi - k)\gy  \phi^2 u_{Rn} -  \phi' u_{Rn}'\right] \cr
&& + \sum_{n={\cal N}_R+1}^{{\cal N}}\frac{2}{a M_P}\left[ (\gy \phi + k)\gy  \phi^2 u_{Rn} -  \phi' u_{Rn}'\right].
\end{eqnarray}
For the gauge field waves with $n \in [1, \,{\cal N}_R]$, the equation of motion is
\begin{equation}
u_{Rn}'' + \left[k^2 - 2 \gy k \phi - \frac{\kappa\Delta{\cal N} }{\gy  M_P^4}(\gy \phi - k)\left(\frac{\phi^2 \phi'}{a^4}\right)'\right]u_{Rn}  
=\frac{2}{a M_P}\left[ a\left(\frac{\phi' }{a}v_R\right)' + (\gy\phi-k) \gy \phi^2 v_R  \right].
\end{equation}
For the gauge field waves with $n \in [{\cal N}_R+1,\,{\cal N}]$, the equation of motion is
\begin{equation}
u_{Rn}'' + \left[k^2 + 2 \gy k \phi + \frac{\kappa\Delta{\cal N} }{\gy  M_P^4}(\gy \phi + k)\left(\frac{\phi^2 \phi'}{a^4}\right)'\right]u_{Rn}  
=\frac{2}{a M_P}\left[ a\left(\frac{\phi' }{a}v_R\right)' + (\gy\phi+k) \gy \phi^2 v_R  \right].
\end{equation}
These latter equations are obtained from the former by replacing $\gy \to -\gy$, recognizing that this also swaps $\Delta{\cal N} \to - \Delta{\cal N}$.
The high-frequency evolution equations similarly become
\begin{eqnarray}
{\cal h}'_R &=& -\frac{1}{a M_P}\left( \sum_{n=1}^{{\cal N}_R}(\phi' + i \gy \phi^2) {\cal y}_{Rn} + \sum_{n={\cal N}_R+1}^{{\cal N}}(\phi' - i \gy \phi^2) {\cal y}_{Rn}\right),  \\
{\cal y}'_{Rn} &=& \frac{1}{a M_P}(\phi' - i \gy \phi^2) {\cal h}_R + i \gy \phi {\cal y}_{Rn} - \frac{i \kappa \Delta{\cal N} }{2 \gy  M_P^4}\left(\phi^2 \phi'/a^4\right)' {\cal y}_{Rn}, \qquad n \in [1, \,{\cal N}_R] \\
{\cal y}'_{Rn} &=& \frac{1}{a M_P}(\phi' + i \gy \phi^2) {\cal h}_R - i \gy \phi {\cal y}_{Rn} - \frac{i \kappa \Delta{\cal N} }{2 \gy  M_P^4}\left(\phi^2 \phi'/a^4\right)' {\cal y}_{Rn}, \qquad n \in [{\cal N}_R+1,\,{\cal N}].
\end{eqnarray}
The equations for the left-handed polarization are obtained by taking $\gy \to -\gy$.

%%%%%%%%%%%%%%%%%%%%%%%%%%%%%%%%%%%%%%%%%%%%%%%%%%%%%%%%%%%
\section{Gravitational waves generated from gauge field perturbations}
\label{app2}

We have focused throughout this article on the behavior of gravitational waves in the presence of an initially smooth gauge field. Here we turn the situation around to investigate the interplay between tensor fluctuations of the gauge field and an initially smooth gravitational field. In particular, we consider the production of gravitational waves in the presence of a superhorizon gauge field wave. For simplicity, the theory is given by following action
\begin{equation}
S = \int d^4 x \sqrt{-g} \left( \frac{1}{2} M_P^2 R - \frac{1}{4}F_{a\mu\nu} F^{a\mu\nu} + {\cal L}_m \right)
\label{eqn:thy}
\end{equation}
All other fields, matter and radiation, are described by ${\cal L}_m$. In this scenario, the equation of motion for $\phi$ is then given by
\begin{equation}
\phi'' + 2\gy^2 \phi^3 = 0
\end{equation}
which is solved analytically in terms of a Jacobi elliptic function
\begin{equation}
\phi = c_1{\rm sn}(c_1(\tau-\tau_i)+c_2 | -1)/\gy
\end{equation}
where $\tau_i$ is some initial time. The constants $c_1$ and $c_2$ are determined at the initial time, which we will take to be deep in the radiation era, and may be expressed in terms of the fraction of critical density in the form of the gauge field, $\Omega_{EDE}$, plus a mixing angle $\theta$ that describes the apportionment of energy between the gauge electric and magnetic fields. Hence, we set
\begin{eqnarray}
\phi_i' &=& a_i^2 H_i M_P \sqrt{2 \Omega_{EDE}} \sin\theta \cr
\phi_i^2 &=&  a_i^2 H_i M_P \sqrt{2 \Omega_{EDE}} \cos\theta/{\gy}
\end{eqnarray}
and allow $\theta \in [0,\, \pi/2]$. We determine that $c_1 = (2 \Omega_{EDE})^{1/4}(\gy H_i M_P)^{1/2} a_i$ and $c_2 = F(\csc^{-1}(\sqrt{\sec\theta}) | -1)$ where $F$ is an elliptic integral of the first kind. Subsequently, we may write
\begin{equation}
\phi = c_1 \psi(x)/\gy,\qquad \phi' = c_1^2 \psi'(x)/\gy
\end{equation}
where $\psi(x) = {\rm sn}(x|-1)$ and $d\psi(x) = {\rm cn}(x|-1){\rm dn}(x|-1)$ and $x = c_1 (\tau - \tau_i) + c_2$.

We again consider tensor fluctuations of the gauge field, and the resulting gravitational waves, as given by Eqs.~(\ref{eqn:Atnsr1}-\ref{eqn:Atnsr2}). We insert these into the action of Eq.~(\ref{eqn:thy}) and expand to second order. In the long-wavelength limit, the equations of motion are identical for both left- and right-circular polarizations so we drop the subscripts, whereby variation with respect to $h$ and $y$ yields
\begin{eqnarray}
&&y'' + 2 \frac{a'}{a}y' + \frac{a''}{a}y = -\frac{2}{a M_P}\left[  \gy^2  \phi^3 h -  \phi' h'\right]
\label{eqn:full1}\\
&&h'' + 2 \frac{a'}{a} h' + \frac{2}{a^2 M_P^2}(\gy^2 \phi^4 - \phi'^2) h = -\frac{2}{a^2 M_P}\left[  -\gy^2  \phi^3 y + \frac{a'}{a}\phi' y + \phi' y'\right]. 
\label{eqn:full2}
\end{eqnarray}
This system of equations can be put in dimensionless form, defining  $\thau = a_i H_i \tau$ and $\tilde g_{Y} = g_{Y} \sqrt{2 \Omega_{YM}} M_P/H_i$ and $H_i$ is the Hubble constant at the starting time, so that $\thau_i=1$:
\begin{eqnarray}
&& \frac{d^2 y}{d\thau^2}+ \frac{2}{\thau}\frac{dy}{d\thau} =-2\frac{\sqrt{2\Omega_{EDE}}}{\thau}\left(\sqrt{\tilde g_{Y}}\psi^3 h - d\psi \frac{dh}{d\thau}\right) 
\label{eqn:simp1}\\
&& \frac{d^2 h}{d\thau^2}+ \frac{2}{\thau}\frac{dh}{d\thau}+ \frac{4 \Omega_{EDE}}{\thau^2}\left(\psi^4 - d\psi^2\right) h  = -2 \frac{\sqrt{2 \Omega_{EDE}}}{\thau}\left(-\sqrt{\tilde g_{Y}}\psi^3 y + d\psi\frac{y}{\thau} + d\psi \frac{dy}{d\thau}\right)
\label{eqn:simp2}
\end{eqnarray}
The argument of the elliptic Jacobi functions is $\psi = \psi(x[\thau])$ where $x = \sqrt{\hat g_{Y}}(\thau-\thau_i)+c_2$.

In previous work we considered the behavior of a primordial gravitational wave spectrum in the presence of an undisturbed flavor-space locked gauge field \cite{Bielefeld:2015daa}. Here, we consider the behavior of the system subject to the initial conditions in which there is no gravitational wave, $h_i = h'_i = 0$, but there is a frozen, superhorizon gauge field wave, $y_i \neq 0$, $y_i'=0$. 

We consider first the case of color electrodynamics, which corresponds to setting $\gy\to 0$ in Eqs.~(\ref{eqn:full1}-\ref{eqn:full2}) or $\psi\to 0$ and $d\psi\to 1$ in Eqs.~(\ref{eqn:simp1}-\ref{eqn:simp2}). The solutions are
\begin{eqnarray}
h &=&y_i \frac{\sqrt{2}}{\alpha}\left[ \sqrt{\frac{1+\alpha}{1-\alpha}}\thau^{-\frac{1}{2}+\frac{\alpha}{2}}
- \sqrt{\frac{1-\alpha}{1+\alpha}}\thau^{-\frac{1}{2}-\frac{\alpha}{2}} - \frac{2 \alpha}{\sqrt{1-\alpha^2}}\right] \cr
y &=& y_i\left[-1 + \frac{2}{\thau}-\frac{2}{\alpha}\thau^{-\frac{1}{2}-\frac{\alpha}{2}} + \frac{2}{\alpha}\thau^{-\frac{1}{2}+\frac{\alpha}{2}}\right]
\end{eqnarray}
where $\alpha =  1 - 16 \Omega_{EDE}$. For $\Omega_{EDE}\ll 1$, the gravitational wave amplitude asymptotes in the future to $-y_i/(2\sqrt{\Omega_{EDE}})$. In this same limit, the gauge field amplitude swaps from $+y_i$ to $-y_i$.

\begin{figure}[ht]
\includegraphics[width=0.45\linewidth]{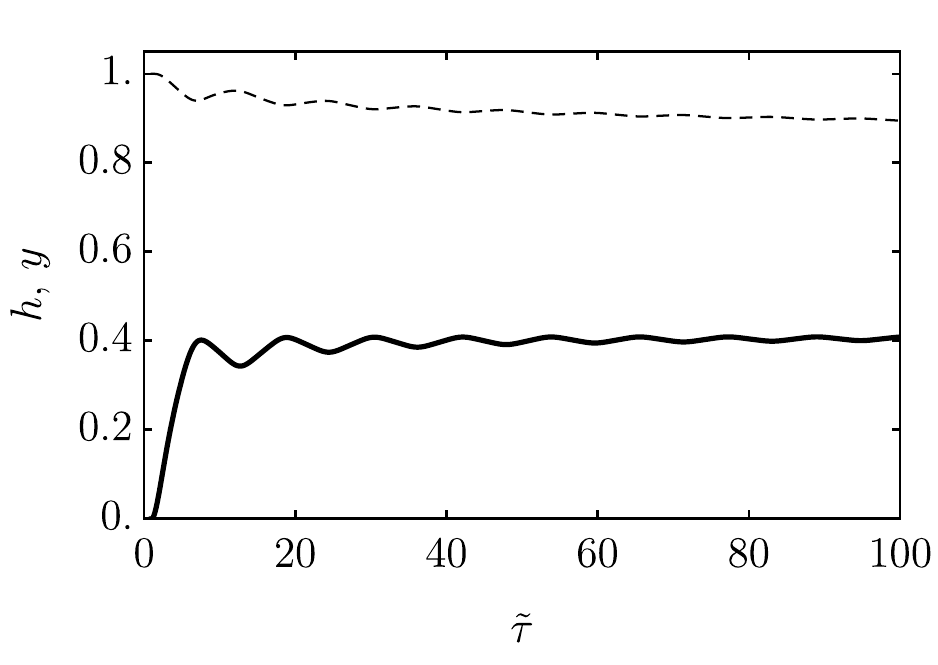}
\caption{A super-horizon gauge field wave creates a super-horizon gravitational wave. The gravitational wave amplitude $h$ (solid) starts at zero but quickly asymptotes to a constant value. The gauge field wave $y$ (dashed) starts at a constant value and decays. In this example we have set $\Omega_{EDE}=0.02$ and $\tilde g_{Y} = 1$. We have chosen initial conditions for the background gauge field $\psi_i=1$ and $d\psi_i=0$.} 
\label{fig:figappb}
\end{figure}

In the case of SU(2), we integrate Eqs.~\ref{eqn:simp1}-\ref{eqn:simp2}, subject to the same initial conditions $h_i = h'_i = 0$, and $y_i \neq 0$, $y_i'=0$. We find that the gravitational wave amplitude grows rapidly and asymptotes to a constant value at time $\thau \gg P/\sqrt{\hat g_{YM}}$ where $P = \Gamma(1/4)^2/\sqrt{2 \pi}$ is the oscillation period of the Jacobi elliptic functions, while the final amplitude depends on the parameters $\gy$ and $\Omega_{YM}$. A sample case is shown in Fig.~\ref{fig:figappb}, which illustrates our main result: a super-horizon gauge field wave creates a super-horizon gravitational wave. We leave for future investigations how this phenomenon may feature in a cosmological scenario.

%%%%%%%%%%%%%%%%%%%%%%%%%%%%%%%%%%%%%%%%%%%%%%%%%%%%%%%%%%%
\bibliography{biblio.bib}

\begin{thebibliography}{39}
\expandafter\ifx\csname natexlab\endcsname\relax\def\natexlab#1{#1}\fi
\expandafter\ifx\csname bibnamefont\endcsname\relax
  \def\bibnamefont#1{#1}\fi
\expandafter\ifx\csname bibfnamefont\endcsname\relax
  \def\bibfnamefont#1{#1}\fi
\expandafter\ifx\csname citenamefont\endcsname\relax
  \def\citenamefont#1{#1}\fi
\expandafter\ifx\csname url\endcsname\relax
  \def\url#1{\texttt{#1}}\fi
\expandafter\ifx\csname urlprefix\endcsname\relax\def\urlprefix{URL }\fi
\providecommand{\bibinfo}[2]{#2}
\providecommand{\eprint}[2][]{\url{#2}}

\bibitem[{\citenamefont{Schutz}(1986)}]{Schutz:1986gp}
\bibinfo{author}{\bibfnamefont{B.~F.} \bibnamefont{Schutz}},
  \bibinfo{journal}{Nature} \textbf{\bibinfo{volume}{323}},
  \bibinfo{pages}{310} (\bibinfo{year}{1986}).

\bibitem[{\citenamefont{Holz and Hughes}(2005)}]{Holz:2005df}
\bibinfo{author}{\bibfnamefont{D.~E.} \bibnamefont{Holz}} \bibnamefont{and}
  \bibinfo{author}{\bibfnamefont{S.~A.} \bibnamefont{Hughes}},
  \bibinfo{journal}{Astrophys. J.} \textbf{\bibinfo{volume}{629}},
  \bibinfo{pages}{15} (\bibinfo{year}{2005}), \eprint{astro-ph/0504616}.

\bibitem[{\citenamefont{Hughes}(2002)}]{Hughes:2001ya}
\bibinfo{author}{\bibfnamefont{S.~A.} \bibnamefont{Hughes}},
  \bibinfo{journal}{Mon. Not. Roy. Astron. Soc.}
  \textbf{\bibinfo{volume}{331}}, \bibinfo{pages}{805} (\bibinfo{year}{2002}),
  \eprint{astro-ph/0108483}.

\bibitem[{\citenamefont{Harry}(2010)}]{Harry:2010zz}
\bibinfo{author}{\bibfnamefont{G.~M.} \bibnamefont{Harry}}
  (\bibinfo{collaboration}{LIGO Scientific}), \bibinfo{journal}{Class. Quant.
  Grav.} \textbf{\bibinfo{volume}{27}}, \bibinfo{pages}{084006}
  (\bibinfo{year}{2010}).

\bibitem[{\citenamefont{Aasi et~al.}(2015)}]{TheLIGOScientific:2014jea}
\bibinfo{author}{\bibfnamefont{J.}~\bibnamefont{Aasi}} \bibnamefont{et~al.}
  (\bibinfo{collaboration}{LIGO Scientific}), \bibinfo{journal}{Class. Quant.
  Grav.} \textbf{\bibinfo{volume}{32}}, \bibinfo{pages}{074001}
  (\bibinfo{year}{2015}), \eprint{1411.4547}.

\bibitem[{\citenamefont{Sathyaprakash et~al.}(2010)\citenamefont{Sathyaprakash,
  Schutz, and Van Den~Broeck}}]{Sathyaprakash:2009xt}
\bibinfo{author}{\bibfnamefont{B.~S.} \bibnamefont{Sathyaprakash}},
  \bibinfo{author}{\bibfnamefont{B.~F.} \bibnamefont{Schutz}},
  \bibnamefont{and} \bibinfo{author}{\bibfnamefont{C.}~\bibnamefont{Van
  Den~Broeck}}, \bibinfo{journal}{Class. Quant. Grav.}
  \textbf{\bibinfo{volume}{27}}, \bibinfo{pages}{215006}
  (\bibinfo{year}{2010}), \eprint{0906.4151}.

\bibitem[{\citenamefont{Punturo et~al.}(2010)}]{Punturo:2010zz}
\bibinfo{author}{\bibfnamefont{M.}~\bibnamefont{Punturo}} \bibnamefont{et~al.},
  \bibinfo{journal}{Class. Quant. Grav.} \textbf{\bibinfo{volume}{27}},
  \bibinfo{pages}{194002} (\bibinfo{year}{2010}).

\bibitem[{\citenamefont{Amaro-Seoane et~al.}(2017)}]{Audley:2017drz}
\bibinfo{author}{\bibfnamefont{P.}~\bibnamefont{Amaro-Seoane}}
  \bibnamefont{et~al.} (\bibinfo{collaboration}{LISA}) (\bibinfo{year}{2017}),
  \eprint{1702.00786}.

\bibitem[{\citenamefont{Caldwell et~al.}(2016)\citenamefont{Caldwell, Devulder,
  and Maksimova}}]{Caldwell:2016sut}
\bibinfo{author}{\bibfnamefont{R.~R.} \bibnamefont{Caldwell}},
  \bibinfo{author}{\bibfnamefont{C.}~\bibnamefont{Devulder}}, \bibnamefont{and}
  \bibinfo{author}{\bibfnamefont{N.~A.} \bibnamefont{Maksimova}},
  \bibinfo{journal}{Phys. Rev.} \textbf{\bibinfo{volume}{D94}},
  \bibinfo{pages}{063005} (\bibinfo{year}{2016}), \eprint{1604.08939}.

\bibitem[{\citenamefont{Maleknejad and
  Sheikh-Jabbari}(2013)}]{Maleknejad:2011jw}
\bibinfo{author}{\bibfnamefont{A.}~\bibnamefont{Maleknejad}} \bibnamefont{and}
  \bibinfo{author}{\bibfnamefont{M.~M.} \bibnamefont{Sheikh-Jabbari}},
  \bibinfo{journal}{Phys. Lett.} \textbf{\bibinfo{volume}{B723}},
  \bibinfo{pages}{224} (\bibinfo{year}{2013}), \eprint{1102.1513}.

\bibitem[{\citenamefont{Mehrabi et~al.}(2017)\citenamefont{Mehrabi, Maleknejad,
  and Kamali}}]{Mehrabi:2015lfa}
\bibinfo{author}{\bibfnamefont{A.}~\bibnamefont{Mehrabi}},
  \bibinfo{author}{\bibfnamefont{A.}~\bibnamefont{Maleknejad}},
  \bibnamefont{and} \bibinfo{author}{\bibfnamefont{V.}~\bibnamefont{Kamali}},
  \bibinfo{journal}{Astrophys. Space Sci.} \textbf{\bibinfo{volume}{362}},
  \bibinfo{pages}{53} (\bibinfo{year}{2017}), \eprint{1510.00838}.

\bibitem[{\citenamefont{Abbott et~al.}(2017{\natexlab{a}})}]{Evans:2016mbw}
\bibinfo{author}{\bibfnamefont{B.~P.} \bibnamefont{Abbott}}
  \bibnamefont{et~al.} (\bibinfo{collaboration}{LIGO Scientific}),
  \bibinfo{journal}{Class. Quant. Grav.} \textbf{\bibinfo{volume}{34}},
  \bibinfo{pages}{044001} (\bibinfo{year}{2017}{\natexlab{a}}),
  \eprint{1607.08697}.

\bibitem[{\citenamefont{Allys et~al.}(2016)\citenamefont{Allys, Peter, and
  Rodriguez}}]{Allys:2016kbq}
\bibinfo{author}{\bibfnamefont{E.}~\bibnamefont{Allys}},
  \bibinfo{author}{\bibfnamefont{P.}~\bibnamefont{Peter}}, \bibnamefont{and}
  \bibinfo{author}{\bibfnamefont{Y.}~\bibnamefont{Rodriguez}},
  \bibinfo{journal}{Phys. Rev.} \textbf{\bibinfo{volume}{D94}},
  \bibinfo{pages}{084041} (\bibinfo{year}{2016}), \eprint{1609.05870}.

\bibitem[{\citenamefont{Rodríguez and Navarro}(2018)}]{Rodriguez:2017wkg}
\bibinfo{author}{\bibfnamefont{Y.}~\bibnamefont{Rodríguez}} \bibnamefont{and}
  \bibinfo{author}{\bibfnamefont{A.~A.} \bibnamefont{Navarro}},
  \bibinfo{journal}{Phys. Dark Univ.} \textbf{\bibinfo{volume}{19}},
  \bibinfo{pages}{129} (\bibinfo{year}{2018}), \eprint{1711.01935}.

\bibitem[{\citenamefont{Adshead and
  Wyman}(2012{\natexlab{a}})}]{Adshead:2012kp}
\bibinfo{author}{\bibfnamefont{P.}~\bibnamefont{Adshead}} \bibnamefont{and}
  \bibinfo{author}{\bibfnamefont{M.}~\bibnamefont{Wyman}},
  \bibinfo{journal}{Phys. Rev. Lett.} \textbf{\bibinfo{volume}{108}},
  \bibinfo{pages}{261302} (\bibinfo{year}{2012}{\natexlab{a}}),
  \eprint{1202.2366}.

\bibitem[{\citenamefont{Adshead and
  Wyman}(2012{\natexlab{b}})}]{Adshead:2012qe}
\bibinfo{author}{\bibfnamefont{P.}~\bibnamefont{Adshead}} \bibnamefont{and}
  \bibinfo{author}{\bibfnamefont{M.}~\bibnamefont{Wyman}},
  \bibinfo{journal}{Phys. Rev.} \textbf{\bibinfo{volume}{D86}},
  \bibinfo{pages}{043530} (\bibinfo{year}{2012}{\natexlab{b}}),
  \eprint{1203.2264}.

\bibitem[{\citenamefont{Bielefeld and
  Caldwell}(2015{\natexlab{a}})}]{Bielefeld:2014nza}
\bibinfo{author}{\bibfnamefont{J.}~\bibnamefont{Bielefeld}} \bibnamefont{and}
  \bibinfo{author}{\bibfnamefont{R.~R.} \bibnamefont{Caldwell}},
  \bibinfo{journal}{Phys. Rev.} \textbf{\bibinfo{volume}{D91}},
  \bibinfo{pages}{123501} (\bibinfo{year}{2015}{\natexlab{a}}),
  \eprint{1412.6104}.

\bibitem[{\citenamefont{Bielefeld and
  Caldwell}(2015{\natexlab{b}})}]{Bielefeld:2015daa}
\bibinfo{author}{\bibfnamefont{J.}~\bibnamefont{Bielefeld}} \bibnamefont{and}
  \bibinfo{author}{\bibfnamefont{R.~R.} \bibnamefont{Caldwell}},
  \bibinfo{journal}{Phys. Rev.} \textbf{\bibinfo{volume}{D91}},
  \bibinfo{pages}{124004} (\bibinfo{year}{2015}{\natexlab{b}}),
  \eprint{1503.05222}.

\bibitem[{\citenamefont{Ade et~al.}(2016)}]{Ade:2015xua}
\bibinfo{author}{\bibfnamefont{P.~A.~R.} \bibnamefont{Ade}}
  \bibnamefont{et~al.} (\bibinfo{collaboration}{Planck}),
  \bibinfo{journal}{Astron. Astrophys.} \textbf{\bibinfo{volume}{594}},
  \bibinfo{pages}{A13} (\bibinfo{year}{2016}), \eprint{1502.01589}.

\bibitem[{\citenamefont{Caldwell and Linder}(2005)}]{Caldwell:2005tm}
\bibinfo{author}{\bibfnamefont{R.~R.} \bibnamefont{Caldwell}} \bibnamefont{and}
  \bibinfo{author}{\bibfnamefont{E.~V.} \bibnamefont{Linder}},
  \bibinfo{journal}{Phys. Rev. Lett.} \textbf{\bibinfo{volume}{95}},
  \bibinfo{pages}{141301} (\bibinfo{year}{2005}), \eprint{astro-ph/0505494}.

\bibitem[{\citenamefont{Doran and Robbers}(2006)}]{Doran:2006kp}
\bibinfo{author}{\bibfnamefont{M.}~\bibnamefont{Doran}} \bibnamefont{and}
  \bibinfo{author}{\bibfnamefont{G.}~\bibnamefont{Robbers}},
  \bibinfo{journal}{JCAP} \textbf{\bibinfo{volume}{0606}}, \bibinfo{pages}{026}
  (\bibinfo{year}{2006}), \eprint{astro-ph/0601544}.

\bibitem[{\citenamefont{Betoule et~al.}(2014)}]{Betoule:2014frx}
\bibinfo{author}{\bibfnamefont{M.}~\bibnamefont{Betoule}} \bibnamefont{et~al.}
  (\bibinfo{collaboration}{SDSS}), \bibinfo{journal}{Astron. Astrophys.}
  \textbf{\bibinfo{volume}{568}}, \bibinfo{pages}{A22} (\bibinfo{year}{2014}),
  \eprint{1401.4064}.

\bibitem[{\citenamefont{Aubourg et~al.}(2015)}]{Aubourg:2014yra}
\bibinfo{author}{\bibfnamefont{E.}~\bibnamefont{Aubourg}} \bibnamefont{et~al.},
  \bibinfo{journal}{Phys. Rev.} \textbf{\bibinfo{volume}{D92}},
  \bibinfo{pages}{123516} (\bibinfo{year}{2015}), \eprint{1411.1074}.

\bibitem[{\citenamefont{Hinshaw et~al.}(2013)}]{Hinshaw:2012aka}
\bibinfo{author}{\bibfnamefont{G.}~\bibnamefont{Hinshaw}} \bibnamefont{et~al.}
  (\bibinfo{collaboration}{WMAP}), \bibinfo{journal}{Astrophys. J. Suppl.}
  \textbf{\bibinfo{volume}{208}}, \bibinfo{pages}{19} (\bibinfo{year}{2013}),
  \eprint{1212.5226}.

\bibitem[{\citenamefont{Ade et~al.}(2014)}]{Ade:2013zuv}
\bibinfo{author}{\bibfnamefont{P.~A.~R.} \bibnamefont{Ade}}
  \bibnamefont{et~al.} (\bibinfo{collaboration}{Planck}),
  \bibinfo{journal}{Astron. Astrophys.} \textbf{\bibinfo{volume}{571}},
  \bibinfo{pages}{A16} (\bibinfo{year}{2014}), \eprint{1303.5076}.

\bibitem[{\citenamefont{Riess et~al.}(2016)}]{Riess:2016jrr}
\bibinfo{author}{\bibfnamefont{A.~G.} \bibnamefont{Riess}}
  \bibnamefont{et~al.}, \bibinfo{journal}{Astrophys. J.}
  \textbf{\bibinfo{volume}{826}}, \bibinfo{pages}{56} (\bibinfo{year}{2016}),
  \eprint{1604.01424}.

\bibitem[{\citenamefont{Di~Valentino et~al.}(2016)\citenamefont{Di~Valentino,
  Melchiorri, and Silk}}]{DiValentino:2016hlg}
\bibinfo{author}{\bibfnamefont{E.}~\bibnamefont{Di~Valentino}},
  \bibinfo{author}{\bibfnamefont{A.}~\bibnamefont{Melchiorri}},
  \bibnamefont{and} \bibinfo{author}{\bibfnamefont{J.}~\bibnamefont{Silk}},
  \bibinfo{journal}{Phys. Lett.} \textbf{\bibinfo{volume}{B761}},
  \bibinfo{pages}{242} (\bibinfo{year}{2016}), \eprint{1606.00634}.

\bibitem[{\citenamefont{Freedman}(2017)}]{Freedman:2017yms}
\bibinfo{author}{\bibfnamefont{W.~L.} \bibnamefont{Freedman}},
  \bibinfo{journal}{Nat. Astron.} \textbf{\bibinfo{volume}{1}},
  \bibinfo{pages}{0121} (\bibinfo{year}{2017}), \eprint{1706.02739}.

\bibitem[{\citenamefont{Abbott
  et~al.}(2017{\natexlab{b}})}]{TheLIGOScientific:2017qsa}
\bibinfo{author}{\bibfnamefont{B.~P.} \bibnamefont{Abbott}}
  \bibnamefont{et~al.} (\bibinfo{collaboration}{LIGO Scientific, Virgo}),
  \bibinfo{journal}{Phys. Rev. Lett.} \textbf{\bibinfo{volume}{119}},
  \bibinfo{pages}{161101} (\bibinfo{year}{2017}{\natexlab{b}}),
  \eprint{1710.05832}.

\bibitem[{\citenamefont{Reitze et~al.}(2019{\natexlab{a}})}]{Reitze:2019dyk}
\bibinfo{author}{\bibfnamefont{D.}~\bibnamefont{Reitze}} \bibnamefont{et~al.},
  \bibinfo{journal}{Bull. Am. Astron. Soc.} \textbf{\bibinfo{volume}{51}},
  \bibinfo{pages}{141} (\bibinfo{year}{2019}{\natexlab{a}}),
  \eprint{1903.04615}.

\bibitem[{\citenamefont{Chen et~al.}(2017)\citenamefont{Chen, Holz, Miller,
  Evans, Vitale, and Creighton}}]{Chen:2017wpg}
\bibinfo{author}{\bibfnamefont{H.-Y.} \bibnamefont{Chen}},
  \bibinfo{author}{\bibfnamefont{D.~E.} \bibnamefont{Holz}},
  \bibinfo{author}{\bibfnamefont{J.}~\bibnamefont{Miller}},
  \bibinfo{author}{\bibfnamefont{M.}~\bibnamefont{Evans}},
  \bibinfo{author}{\bibfnamefont{S.}~\bibnamefont{Vitale}}, \bibnamefont{and}
  \bibinfo{author}{\bibfnamefont{J.}~\bibnamefont{Creighton}}
  (\bibinfo{year}{2017}), \eprint{1709.08079}.

\bibitem[{\citenamefont{Reitze et~al.}(2019{\natexlab{b}})}]{Reitze:2019iox}
\bibinfo{author}{\bibfnamefont{D.}~\bibnamefont{Reitze}} \bibnamefont{et~al.},
  \bibinfo{journal}{Bull. Am. Astron. Soc.} \textbf{\bibinfo{volume}{51}},
  \bibinfo{pages}{035} (\bibinfo{year}{2019}{\natexlab{b}}),
  \eprint{1907.04833}.

\bibitem[{\citenamefont{Mills et~al.}(2018)\citenamefont{Mills, Tiwari, and
  Fairhurst}}]{Mills:2017urp}
\bibinfo{author}{\bibfnamefont{C.}~\bibnamefont{Mills}},
  \bibinfo{author}{\bibfnamefont{V.}~\bibnamefont{Tiwari}}, \bibnamefont{and}
  \bibinfo{author}{\bibfnamefont{S.}~\bibnamefont{Fairhurst}},
  \bibinfo{journal}{Phys. Rev.} \textbf{\bibinfo{volume}{D97}},
  \bibinfo{pages}{104064} (\bibinfo{year}{2018}), \eprint{1708.00806}.

\bibitem[{\citenamefont{Hirata et~al.}(2010)\citenamefont{Hirata, Holz, and
  Cutler}}]{Hirata:2010ba}
\bibinfo{author}{\bibfnamefont{C.~M.} \bibnamefont{Hirata}},
  \bibinfo{author}{\bibfnamefont{D.~E.} \bibnamefont{Holz}}, \bibnamefont{and}
  \bibinfo{author}{\bibfnamefont{C.}~\bibnamefont{Cutler}},
  \bibinfo{journal}{Phys. Rev.} \textbf{\bibinfo{volume}{D81}},
  \bibinfo{pages}{124046} (\bibinfo{year}{2010}), \eprint{1004.3988}.

\bibitem[{\citenamefont{Tamanini et~al.}(2016)\citenamefont{Tamanini, Caprini,
  Barausse, Sesana, Klein, and Petiteau}}]{Tamanini:2016zlh}
\bibinfo{author}{\bibfnamefont{N.}~\bibnamefont{Tamanini}},
  \bibinfo{author}{\bibfnamefont{C.}~\bibnamefont{Caprini}},
  \bibinfo{author}{\bibfnamefont{E.}~\bibnamefont{Barausse}},
  \bibinfo{author}{\bibfnamefont{A.}~\bibnamefont{Sesana}},
  \bibinfo{author}{\bibfnamefont{A.}~\bibnamefont{Klein}}, \bibnamefont{and}
  \bibinfo{author}{\bibfnamefont{A.}~\bibnamefont{Petiteau}},
  \bibinfo{journal}{JCAP} \textbf{\bibinfo{volume}{1604}}, \bibinfo{pages}{002}
  (\bibinfo{year}{2016}), \eprint{1601.07112}.

\bibitem[{\citenamefont{Soares-Santos et~al.}(2019)}]{Soares-Santos:2019irc}
\bibinfo{author}{\bibfnamefont{M.}~\bibnamefont{Soares-Santos}}
  \bibnamefont{et~al.} (\bibinfo{collaboration}{DES, LIGO Scientific, Virgo}),
  \bibinfo{journal}{Astrophys. J.} \textbf{\bibinfo{volume}{876}},
  \bibinfo{pages}{L7} (\bibinfo{year}{2019}), \eprint{1901.01540}.

\bibitem[{\citenamefont{Farr et~al.}(2019)\citenamefont{Farr, Fishbach, Ye, and
  Holz}}]{Farr:2019twy}
\bibinfo{author}{\bibfnamefont{W.~M.} \bibnamefont{Farr}},
  \bibinfo{author}{\bibfnamefont{M.}~\bibnamefont{Fishbach}},
  \bibinfo{author}{\bibfnamefont{J.}~\bibnamefont{Ye}}, \bibnamefont{and}
  \bibinfo{author}{\bibfnamefont{D.}~\bibnamefont{Holz}},
  \bibinfo{journal}{Astrophys. J.} \textbf{\bibinfo{volume}{883}},
  \bibinfo{pages}{L42} (\bibinfo{year}{2019}), \eprint{1908.09084}.

\bibitem[{\citenamefont{Zhao et~al.}(2011)\citenamefont{Zhao, Van Den~Broeck,
  Baskaran, and Li}}]{Zhao:2010sz}
\bibinfo{author}{\bibfnamefont{W.}~\bibnamefont{Zhao}},
  \bibinfo{author}{\bibfnamefont{C.}~\bibnamefont{Van Den~Broeck}},
  \bibinfo{author}{\bibfnamefont{D.}~\bibnamefont{Baskaran}}, \bibnamefont{and}
  \bibinfo{author}{\bibfnamefont{T.~G.~F.} \bibnamefont{Li}},
  \bibinfo{journal}{Phys. Rev.} \textbf{\bibinfo{volume}{D83}},
  \bibinfo{pages}{023005} (\bibinfo{year}{2011}), \eprint{1009.0206}.

\bibitem[{\citenamefont{Cai and Yang}(2017)}]{Cai:2016sby}
\bibinfo{author}{\bibfnamefont{R.-G.} \bibnamefont{Cai}} \bibnamefont{and}
  \bibinfo{author}{\bibfnamefont{T.}~\bibnamefont{Yang}},
  \bibinfo{journal}{Phys. Rev.} \textbf{\bibinfo{volume}{D95}},
  \bibinfo{pages}{044024} (\bibinfo{year}{2017}), \eprint{1608.08008}.

\end{thebibliography}
%%%%%%%%%%%%%%%%%%%%%%%%%%%%%%%%%%%%%%%%%%%%%%%%%%%%%%%%%%%
\end{document}